\documentclass[a4paper,11pt]{article}
\pdfoutput=1 
\usepackage{jcappub} 
\usepackage[T1]{fontenc} 
\usepackage[utf8]{inputenc}
\usepackage{xspace}
\usepackage{siunitx}
\usepackage{amsmath,amssymb} 
\usepackage{xcolor}
\usepackage{siunitx}
\usepackage{booktabs}
\usepackage{multirow}

\newcommand{\pythia}{\textsf{Pythia}\xspace}
\newcommand{\pythias}{\textsf{Pythia6}\xspace}
\newcommand{\pythiae}{\textsf{Pythia8}\xspace}
\newcommand{\herwig}{\textsf{Herwig}\xspace}
\newcommand{\herwigs}{\textsf{Herwig7}\xspace}
\newcommand{\madgraph}{\textsf{MadGraph5}\xspace}
\newcommand{\madevent}{\textsf{MadEvent}\xspace}
\newcommand{\feynrules}{\textsf{FeynRules}\xspace}
\newcommand{\wimpsim}{\textsf{WimpSim}\xspace}
\newcommand{\darksusy}{\textsf{DarkSUSY}\xspace}
\newcommand{\hs}{\textsf{H7}\xspace}
\newcommand{\pe}{\textsf{P8}\xspace}
\newcommand{\ps}{\textsf{P6}\xspace}
\newcommand{\pppc}{\textsf{PPPC}\xspace}
\newcommand{\ekin}{\ensuremath E_{\rm{kin}}}

\DeclareSIUnit{\gev}{\giga\electronvolt}
\DeclareSIUnit{\tev}{\tera\electronvolt}
\DeclareSIUnit{\au}{AU}

\title{Effect of polarisation and choice of event generator on spectra from dark matter annihilations}

\author[a]{Carl Niblaeus,}
\author[b]{Jonathan M. Cornell,}
\author[a]{Joakim Edsj\"o}
\affiliation[a]{Stockholm University \& the Oskar Klein Centre, Stockholm, Sweden}
\affiliation[b]{Department of Physics, University of Cincinnati, Cincinnati, OH 45221, USA}

\emailAdd{carl.niblaeus@fysik.su.se}
\emailAdd{jonathan.cornell@uc.edu}
\emailAdd{edsjo@fysik.su.se}

\abstract{If indirect detection searches are to be used to discriminate between dark matter particle models, it is crucial to understand the expected energy spectra of secondary particles such as neutrinos, charged antiparticles and gamma rays emerging from dark matter annihilations in the local Universe. In this work we study the effect that both the choice of event generator and the polarisation of the final state particles can have on these predictions. For a variety of annihilation channels and dark matter masses, we compare yields obtained with \pythiae and \herwigs of all of the aforementioned secondary particle species. We investigate how polarised final states can change these results and do an extensive study of how the polarisation can impact the expected flux of neutrinos from dark matter annihilations in the centre of the Sun.

We find that differences between the event generators are larger for yields of hadronic end products such as antiprotons, than for leptonic end products. Concerning polarisation, we conversely find the largest differences in the leptonic spectra. The large differences in the leptonic spectra point to the importance of including polarisation effects in searches for neutrinos from dark matter annihilations in the Sun. However, we find that these differences are ultimately somewhat washed out by propagation effects of the neutrinos in the Sun.}

\begin{document}
\maketitle
\flushbottom

\section{Introduction} \label{sec:intro}
Despite years of investigation, the identity of dark matter remains an open question. One of the most widely studied ideas is that dark matter consists of a hitherto unknown type of particle called a WIMP (Weakly Interacting Massive Particle), which is well motivated because it allows for the production of the correct abundance of dark matter through a simple thermal mechanism in the early universe \cite{Bertone:2004pz}. The same annihilation processes that are important in setting the WIMP relic abundance also lead to substantial astronomical fluxes of Standard Model (SM) particles from regions of space where the dark matter density is high. This leads to the programme of so-called indirect detection, where one attempts to discover dark matter particles, such as WIMPs, by observing excesses of these particles, typically charged antiparticles, gamma-rays or neutrinos, compared to the astrophysical background.

To be able to predict the signal expected in indirect detection for a particular dark matter model it is essential to be able to accurately predict the energy spectra of the yield particles (by yield particles we here refer to the particles one looks for in indirect detection experiments) expected directly from annihilations for a given dark matter model. To compare with observations, these production yields are translated to an expected flux at Earth by propagating the produced SM particles from the source to the Earth, including necessary effects such as diffusion by magnetic fields in the case of charged particles, or flavour oscillations in the case of neutrinos. Depending on the astrophysical environment where the dark matter annihilation takes place, other effects can also play an important role, for example accounting for the interactions of neutrinos with the solar material is crucial in the searches for neutrinos from dark matter annihilations in the Sun \cite{Blennow:2007tw,Baratella:2013fya}. The characteristics of the production yields are therefore a vital part of the indirect detection programme, and any uncertainty in the production yields will translate into an uncertainty in experimental constraints on dark matter properties. 

To determine the production yields, one often uses a model-independent approach where the specific particle physics model containing the dark matter particle (\textit{e.g.} the neutralino in a supersymmetric model \cite{Martin:1997ns}) is not specified, but rather Monte Carlo simulations are done for various two body final states at a range of  centre-of-mass energies. By interpolating between the tabulated yields from these simulations, the energy spectra of the yield particles of interest can be determined for annihilations of a dark matter particle of arbitrary mass to a particular final state (it is usually assumed $E_{\rm c.o.m.} = 2 m_\chi$, where $m_\chi$ is the dark matter mass). The total injection rate at the source $\mathcal{Q}$ of a particular yield particle $f$ is then given by the relation 
\begin{equation}
\frac{d\mathcal{Q}}{dE_f} = \frac{1}{N_\chi}\frac{\rho_\chi^2}{m_\chi^2}\sum_i \langle \sigma_i v \rangle \frac{dN_i}{dE_f}\, ,
\end{equation}
where $\langle \sigma_i v \rangle$  is the thermally averaged annihilation cross section for annihilation to final state $i$, $N_\chi = 2$ for self-conjugate dark matter and 4 otherwise, and $dN_i/dE_f$ is the corresponding spectra of yield particle $f$ for the that final state, as determined from the aforementioned interpolation. For parameter space scans, using this approach rather than than running Monte Carlo simulations for every model point leads to overwhelming speed increases. However, there are some models in which this procedure can fail to capture important physics. For example, internal bremsstrahlung, where a gauge boson is emitted from one of the internal propagators of the annihilation process, can change significantly the shape of the obtained photon spectrum \cite{Bringmann:2007nk,Bringmann:2015cpa}. Also, as the final state particles are often assumed to be unpolarised in this approach, it can lead to substantial mismodelling of the yields in models in which the final state particle are produced with a particular polarisation \cite{Barger:2007xf,Cirelli:2008pk,Tang:2015meg,Garcia-Cely:2016pse,Buckley:2013sca,Ellis:2017tkh}. 

One typically employs general purpose Monte Carlo event generators \cite{Sjostrand:2014zea,Bahr:2008pv,Bellm:2015jjp,Bellm:2017bvx,Buckley:2011ms} to simulate the dark matter annihilations and the resulting energy spectra of yield particles in order to handle the complications of parton showering and hadronisation in the case of coloured final state particles. These event generators include the treatment of parton showering and hadronisation and the subsequent decays into stable particles. Apart from coloured radiation also electromagnetic and sometimes electroweak radiation can be included \cite{Christiansen:2014kba}. There are however differences in how these codes treat the simulation of the involved particle physics. In particular, since hadronisation is a non-perturbative process, one is forced to resort to phenomenological modelling for which the codes follow varying approaches. Also the radiation of coloured particles, the parton showering, can be different from code to code. Therefore, it is important to quantify the differences between the different event generators in order to assess their ultimate impact on indirect dark matter searches.

In this paper we look at two of the most common event generators used for showering and hadronisation, \pythiae \cite{Sjostrand:2006za,Sjostrand:2014zea} and \herwigs \cite{Bellm:2015jjp,Bahr:2008pv}, and look at the differences between them for the energy spectra of the various yield particles of interest in indirect detection experiments. We also compare to results obtained  previously with \pythias \cite{Sjostrand:2006za}. These spectra are implemented in the current version of the \darksusy code \cite{Bringmann:2018lay} and are thus useful as a reference. In addition to comparing the yields between event generators, we consider the effect of spin polarisation on the spectra, looking specifically at how this affects the predictions for event rates in neutrino telescopes, where these effects are expected to have a significant impact. 

There are a number of previous studies that consider the yields from dark matter annihilations. In refs.~\cite{Cirelli:2010xx,Baratella:2013fya} in particular, a large number of annihilation channels, dark matter masses and yield particles are considered, with the results publicly available. In this paper, we perform a complementary analysis while also updating refs.~\cite{Cirelli:2010xx,Baratella:2013fya} with results from new versions of the event generators and including further details on the effects of final state polarisation. The effect of varying the event generator and its parameters on the gamma-ray flux from dark matter annihilations has been studied in refs.~\cite{Cirelli:2010xx, Cembranos:2013cfa, Amoroso:2018qga}. Of these studies, ours is similar in spirit to refs.~\cite{Cirelli:2010xx, Cembranos:2013cfa}, but we expand upon these works by considering a broader range of yield particles beyond photons, as well as polarised final states. The approach of ref.~\cite{Amoroso:2018qga} is different. There the authors focus mainly on the variance in the photon yield that is derived from the uncertainties of the QCD model parameters in \pythia, while we (and the previously mentioned studies) focus on comparing the results of various event generators with their default tunes. Even with this less comprehensive approach, we still find significant uncertainties in the predicted fluxes of all of the yield particles of interest, hopefully motivating detailed studies along the lines of ref.~\cite{Amoroso:2018qga} that consider yield particles other than photons, as well as the effects of varying the tune parameters in \herwig as well as \pythia. Finally, the effect of polarisation in gauge boson final states in dark matter annihilations in the Sun has been studied in ref.~\cite{Jungman:1994jr}, where a 5-10\% effect is found on the second moment of the neutrino fluxes. For neutralino dark matter, the effect of polarisation on gauge boson final states was also studied in ref.~\cite{Barger:2007xf} where they emphasise the importance of including the polarisation effects. Here, we study polarisation of gauge bosons and also extend the study to fermion polarisations and re-investigate the effect of polarisation in a more complete framework for the simulation of the neutrino propagation from the centre of the Sun to the Earth. 

\section{Final state polarisation in dark matter annihilations} \label{sec:pol}
In many cases, the final states in the dark matter annihilation processes can be preferentially produced in a specific helicity configuration. This will affect the shapes of the spectra of the yield particles \cite{Garcia-Cely:2016pse,Ibarra:2016fco}, a result of non-isotropic angular distributions of the particles that emerge from the decay of the polarised particle. Consider for example the annihilation of dark matter particles into a polarised pair of final state particles. If the final state particles decay through a two-body decay, the emitted particles are monochromatic in the rest frame of the decaying particle. Their angular decay distributions are however not necessarily isotropic, with some directions preferred over others in the decay, for example when the decay proceeds through a chiral coupling. Therefore, when the decay products are boosted to the rest frame of the annihilation, they will have an energy distribution which is generally non-flat. The flat distribution arises only in case the final state particle decays isotropically (as for example for a decaying scalar) or is unpolarised.

In the case of annihilation into massive gauge bosons, some degree of polarisation is expected from the separation of the longitudinal and transverse degrees of freedom---the longitudinal degrees of freedom are connected to the Goldstone bosons of electroweak symmetry breaking and thus connected to the interactions with the Higgs sector, whereas the transverse degrees of freedom are associated with the pure gauge part \cite{Drees:1992am,Garcia-Cely:2016pse}. Depending on the strength of the interaction of the dark sector with the Higgs and pure gauge part respectively, the production rate of longitudinal and transverse polarised states can therefore differ. As an example, we note that, in neutralino annihilation into a pair of gauge bosons, the gauge bosons can be produced with purely transverse polarisation \cite{Drees:1992am,Barger:2007xf}.

Another case where polarisation can play an important role is when a Dirac dark matter candidate annihilates with its conjugate particle to SM fermions in a spin 1 configuration through a vector mediator. It is possible for the SM fermions to be produced with a net polarisation if this vector mediator $B^\mu$ has a chiral coupling to the SM fermions, \textit{i.e.} if it couples to the fermions via a term of the form $B^\mu (g_L \bar f \gamma_\mu P_L f + g_R \bar f \gamma_\mu P_R f)$. If $g_R > g_L$, dark matter annihilations to left-handed SM fermions and right-handed anti-fermions will dominate (and vice versa). Such a coupling structure can arise in $Z'$ models that are anomaly free \cite{Ellis:2017tkh}. Dirac neutralino dark matter can also lead to a similar excess of one polarisation state over another \cite{Buckley:2013sca}.

In general, the polarisation of the final state particles depends on the nature of the interactions between the dark sector and the SM. It is therefore difficult to make model-independent statements about what polarisation to expect. Here we refrain from committing to a specific particle physics model and see what effect the polarisation can have by considering spectra from pure final state polarisations that should in most cases bracket the spectra in a full particle physics model. 

\section{Event generators} \label{sec:evtgen}
An event generator uses Monte Carlo methods to simulate the  complexity of a particle physics process \cite{Buckley:2011ms, Beringer:1900zz}. In order to obtain fluxes of the stable particles from a particular dark matter annihilation process we have used two of the more popular general purpose event generators, \pythiae and \herwigs, for hadronisation and showering of hard process events simulated with \madgraph. For a given final state of the annihilation process, the event generator is responsible for the evolution of the hard process final states into the final list of stable particles from the annihilation. This includes decay of unstable annihilation products, radiation of coloured particles and photons as well as the hadronisation of partons and the subsequent decay of unstable hadrons. We will in the following review the main characteristics of general purpose event generators, restricting ourselves to that which is relevant for the dark matter annihilation fluxes. We then specifically consider the properties of \pythiae and \herwigs. 

\subsection{General purpose event generators}
Event generators typically divide the modelling of a particle physics process into several parts. The high momentum transfer interaction is called the hard process. If coloured particles are produced in the hard process, they undergo a showering process,
successively radiating coloured particles and evolving downwards to lower scale until they reach a cutoff $Q_0$, typically around \SI{1}{\gev}, which signifies the transition to the non-perturbative regime of QCD. At this point, all partons form colour-neutral hadrons in a process called hadronisation. Unstable hadrons are then decayed, leading to the ultimate list of stable particles that are produced in the interaction.\footnote{Note that whether a particle is defined as stable or not depends on the application, in an accelerator all particles that decay outside the detector are denoted stable. In the current situation where we consider dark matter annihilations in an astrophysical environment, we have to make sure that these particles are allowed to decay.} Non-coloured particles are decayed to stable states as well, and photon and electroweak radiation can be taken into account in the showering and decay processes.\footnote{In general one also has to consider radiation of the incoming partons in the hard process as well as multiple interactions between incoming partons for collisions of composite particles; this is however not necessary in the current case of dark matter annihilations, since the dark matter particle is assumed to be an uncharged elementary particle.}

The hard process is represented by the matrix element of the process in question; in the case of annihilation into coloured states it is the partonic amplitude. In this paper we take the hard process to be dark matter annihilating via an $s$-channel mediator to a pair of SM particles. While our results are broadly applicable to a range of dark matter models considered in the literature, it is important to note that there exist models in which final states other than two SM particles dominate, such as secluded dark matter scenarios (\textit{e.g.} \cite{Pospelov:2007mp, Elor:2015bho}) or models in which radiation off a charged $t$-channel mediator is important (\textit{e.g.} \cite{Bringmann:2007nk, Kopp:2014tsa}).

In the case of partonic annihilation channels, or when the annihilation channel particle decays to quarks, the outgoing partons from the hard process radiate coloured particles, losing energy in the process. These parton showers take into account radiation of coloured particles like quarks and gluons, and also photons and electroweak bosons \cite{Christiansen:2014kba} can be included in the showering procedure. This radiation is dominated by collinear and soft emission and therefore showering algorithms have traditionally focused on radiation in this limit. The showering is performed iteratively, starting from the outgoing coloured states from the hard process and evolving down to a scale $\sim \SI{1}{\gev}$, where hadronisation kicks in. 

The choice of evolution variable in the showers differs between generators. Older \pythia versions evolved in terms of the virtuality of the partons \cite{Sjostrand:2006za}. Using the virtuality as evolution variable can be shown to be incorrect when moving away from the collinear and soft limit, due to insufficient modelling of  colour coherence effects \cite{Gustafson:1986db}. To take these effects into account one can instead use an angular ordering, evolving in the angle between the outgoing partons in the shower \cite{Marchesini:1987cf, Marchesini:1989yk}. Another way of accounting for the colour coherence effects is to consider radiation from colour dipoles instead of partons, so that dipole branchings are considered rather than parton branchings \cite{Gustafson:1986db,Gustafson:1987rq,Lonnblad:1992tz}.

Hadronisation and hadron decays is the final part of the event generation. This consists of the formation of colour neutral hadrons out of the partons coming from the parton showering. Since this occurs in the non-perturbative regime of QCD one has to resort to phenomenological modelling of the hadronisation process. The two most common models are the Lund string model \cite{Andersson:1983ia,Sjostrand:1984ic} and the clustering model \cite{Webber:1983if}. The hadronisation part of the event generation includes  phenomenological parameters that must be tuned to experimental data. 

In the string model, all the partons from the showering are connected by strings that represent the strong force potential between them. These strings are stretched as partons continue to move outwards and splits when the potential energy is sufficient to create a new $q \bar{q}$ pair. The gluons in the shower are represented by kinks on the strings. Further complexities allow for baryon formation. After string evolution stops, unstable hadrons are decayed.

The clustering model starts by enforcing a splitting of all gluons into $q\bar{q}$ pairs. Colour neutral clusters are then formed out of the resulting quarks and antiquarks with the clusters then forming low-mass hadrons, where the specific hadron is determined mainly by phase space and mass. Any unstable hadrons are then decayed.

\subsection{Comparing \pythiae and \herwigs}
The history of \pythia goes back to the development of the Lund string model \cite{Sjostrand:1982fn}. Today it is one of the most used event generators. The latest major release \pythiae \cite{Sjostrand:2014zea} consisted, apart from physics updates, of a complete rewrite of the old \textsf{Fortran} version \pythias \cite{Sjostrand:2006za} into a \textsf{C++} version. 

For the parton showering, \pythiae uses a modified version of the dipole showering scheme, where a variable closely related to the transverse momentum is used as the evolution variable \cite{Sjostrand:2004ef,Corke:2010yf}. The hadronisation is performed with the Lund string model. 

\herwigs \cite{Bellm:2017bvx, Bellm:2015jjp,Bahr:2008pv} continues the development of \textsf{Herwig++} \cite{Bahr:2008pv} which in turn was a major \textsf{C++} rewrite of the the earlier \textsf{Fortran} version. \herwig has traditionally relied on angular-ordered parton showers, this continues to be the case in \herwigs, with additional options for dipole showering also included. Hadronisation is in \herwigs performed using the clustering model, as in older versions. 

By using helicity amplitudes, \herwigs tracks spin throughout the event generation machinery and therefore generally includes spin correlation effects. This is generally not the case in \pythiae where the effects are included only for some specific cases. As described in section~\ref{sec:pol}, such effects can significantly change the shape of the energy distributions. 

Given the differences between \pythiae and \herwigs in the modelling of showers and hadronisation, we expect some differences in the fluxes of the final yield particles, even though the generators are for the most part tuned to the same data sets. This is particularly true in the case of annihilation into quarks or for the fluxes of, for example, anti-protons where QCD effects are expected to be more important. 

\section{Simulation procedure} \label{sec:simulation}
To obtain the fluxes of the yield particles of interest we have simulated annihilations of dark matter particles at rest into an $s$-channel uncharged resonance of mass $2m_\chi$ that decays into the desired annihilation channel. We then let the annihilation products undergo parton showers, decays, and hadronisation. All unstable particles (including muons, pions, kaons and neutrons except for the case of annihilations inside the Sun) are forced to decay. In each annihilation we count the number of yield particles appearing and record each particle's kinetic energy. Dividing by the number of annihilations generated we thus obtain $dN/d\ekin$---the number of yield particles per annihilation as a function of their kinetic energy.

We have considered a mix of quark, lepton, and gauge boson annihilation channels
\begin{align}
\tau_L^-\tau^+_R,\ \tau_R^-\tau^+_L,\ t_L\overline{t}_R,\ t_R\overline{t}_L,\ b\overline{b},\ W^-_L W^+_L,\ W^-_T W^+_T
\end{align}
with the heavier quark and leptons chosen because they will lead to more complicated showering and decay processes. Polarised $b$ quarks are not considered since these hadronise before decaying and the polarisation is lost in the showering and hadronisation. The missing $Z^0Z^0$ final state has similar phenomenology to the $W^+W^-$, with the important difference that there will be no QED bremsstrahlung off the $Z$.\footnote{Note that our choice of polarisations for the final state fermions differs from that of the \pppc \cite{Cirelli:2010xx}, which instead includes final states in which both the fermion and anti-fermion are in the same polarisation state. The reason behind this is theoretical -- while a chiral coupling of the fermions to a vector boson of the form discussed in section~\ref{sec:pol} can lead to annihilations to fermions in a definite LR or RL helicity state, the need for the Lagrangian to be Hermitian seems to exclude the possibility of preferential annihilations to LL or RR final states. For example two dark matter particles in a $J=0$ configuration could annihilate to two fermions in a RR state if they annihilated via an $s$-channel scalar mediator $\phi$ with couplings of the form $\phi \bar f P_L f$. However such a term is not Hermitian by itself, so its Hermitian conjugate, $\phi \bar f P_R f$, must also be included in the Lagrangian, leading to final state fermions that are unpolarised on average. We also do not directly simulate gauge boson final states that are a mix of transverse and longitudinal polarisations. Assuming that $W^-_L W^+_T$ is produced at the same rate as $W^-_T W^+_L$, the average spectra from an annihilation to these states can be determined by combining our results via the relation $(W^-_L W^+_T + W^-_T W^+_L)/2 = (W^-_L W^+_L + W^-_T W^+_T)/2$.} For these final states, we have constructed the energy distributions for the yield particles $e^\pm$, $p$, $\bar{p}$, $\gamma$ and all $\nu_\ell$, $\bar{\nu}_\ell$ for $\ell=e, \mu, \tau$. For the annihilation channels, the unpolarised state is used in the cases where there is no subscript, otherwise the subscript indicates the helicities of the particles. The subscript denotes for fermions whether it is right- ($R$) or left-handed ($L$), while for gauge bosons a subscript of $L$ ($T$) indicates longitudinal (transverse) polarisation, with the two transverse polarisation states assumed to occur at equal rates.\footnote{To convert our results to the yields from unpolarised final states, the simple relations $f^-f^+ = (f^-_L f^+_R + f^-_R f^+_L)/2$ and $W^- W^+ = (2 W^-_T W^+_T + W^-_L W^+_L)/3$ can be used.}
As described below, we have used \madgraph and \madevent \cite{Alwall:2014hca} for generation of hard process events and some decays, while  \pythiae or  \herwigs are used to simulate the subsequent showering, hadronisation, and other decays not handled by \madevent. We use \madgraph rather than the internal \pythia matrix elements because of the limited ability of \pythia to handle polarised final state particles. The same hard process events from \madevent are then used in \herwig for consistency in our comparisons. We have used version 8.240 of \pythia, version 2.6.5 of \madgraph  and version 7.1.1 of \herwig in our simulations. These will be referred to as \pythiae, \madgraph and \herwigs throughout. 

For the \madgraph/\madevent simulations, we have set up a simple \feynrules model that includes a scalar resonance (for the gauge boson channels) and a spin-1 resonance (for the fermionic channels), which couple to all fermionic annihilation channel particles through maximally parity violating couplings with left- and right-handed projectors.\footnote{A vector or pseudo-vector resonance that couples to the fermions via $\bar f \gamma_\mu f$ and $\bar f \gamma_\mu  \gamma_5 f$ respectively would give  unpolarised final states on average.} We can then control the polarisation of fermion final states by setting these couplings. For the gauge boson final states we achieve a longitudinally polarised final state by coupling the resonance to a term of the form $V_\mu V^\mu$, for gauge fields $V^\mu$, whereas the transverse polarisation is achieved by a coupling to the squared field strength tensor of the gauge field, $F^{\mu\nu}{F}_{\mu\nu}$. A pseudoscalar mediator which couples to $F^{\mu\nu}\widetilde{F}_{\mu\nu}$ would result in the same energy spectrum \cite{Cirelli:2008pk}.  We implement the model in \madgraph via the UFO  \cite{Degrande:2011ua} interface, allowing us to use it in event generation. From the \madevent simulations, we produce event files in the Les Houches Event File (LHEF) format \cite{Alwall:2006yp,Andersen:2014efa}, which are then read into \pythiae and \herwigs to perform the showering and hadronisation.\footnote{In \herwigs, the \textsf{EvtGen} code is used for the decay of heavy hadrons containing $b$- or $c$-quarks. In order to improve efficiency in the simulations, we have had to turn off this interface and use the internal \herwigs decays instead. This affects primarily the $b\overline{b}$ channel, where many of these heavy hadrons are produced. The difference is generally smaller than the difference compared to \pythiae simulations.} 

For annihilation channels into unstable particles where we are interested in polarisation effects, and there are no built-in features in the event generator to take into account the polarisation of the final state, we let \madevent perform the decays. We use the decay chain syntax to decay electroweak gauge bosons and top quarks down to parton or lepton level. The top quarks are decayed as $t\to b W^+$ and electroweak gauge bosons are decayed into all possible two-body decays. Since there are functionalities in \pythiae and \herwigs to perform polarised $\tau$ decays \cite{Ilten:2012zb,Grellscheid:2007tt}, we do not decay these in \madgraph but use \pythiae or \herwigs. Finally, following the code defaults and in order to make consistent comparisons we do not turn on electroweak radiation in the showers (this is turned off by default in \pythiae and not available in \herwigs). 

\section{Results for annihilation fluxes} \label{sec:results}
In this section we look at the fluxes of the various yield particles of interest produced in the annihilations. Broadly speaking, for a given annihilation channel, the flux of a specific yield particle will be harder the earlier it can appear in the decay chain, with the hardest flux produced when the dark matter particles annihilate directly into the yield particle in question. Typically, there is also a soft component of the spectrum from radiation and decays of lower energy particles produced in the hadronisation. In the following, we go into more details on the spectra obtained for the various annihilation channels.

\subsection{Differences between \pythiae and \herwigs}
\begin{figure}[t]
    \centering
    \includegraphics[height=0.3\textheight,clip,trim={0 0 0 8mm}]{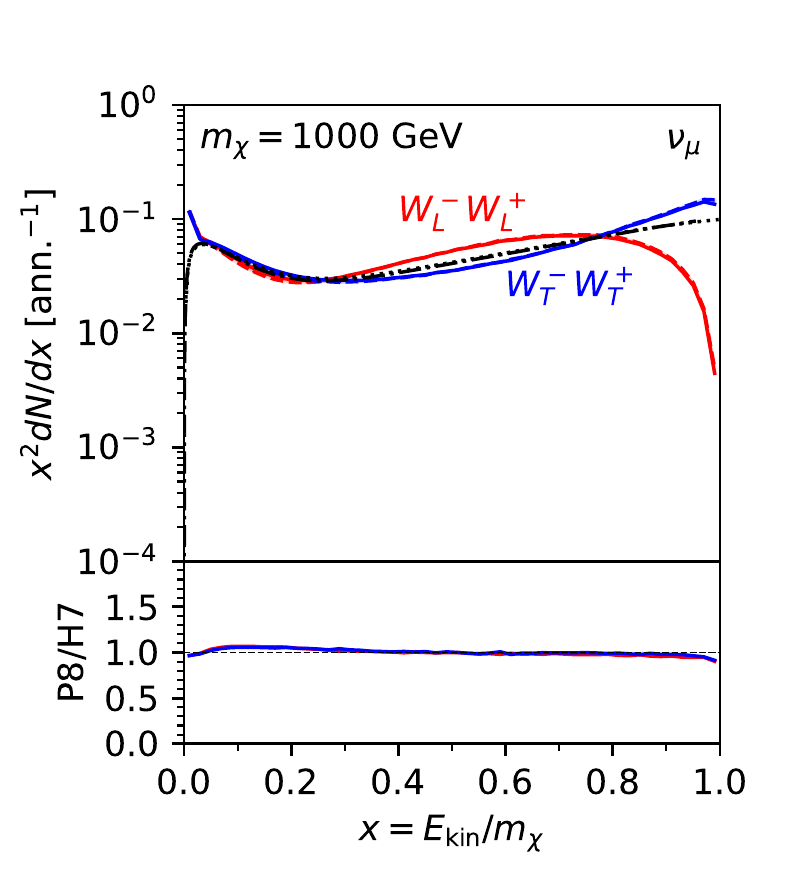}\hspace{-1.0em}
    \includegraphics[height=0.3\textheight,clip,trim={14mm 0 0 8mm}]{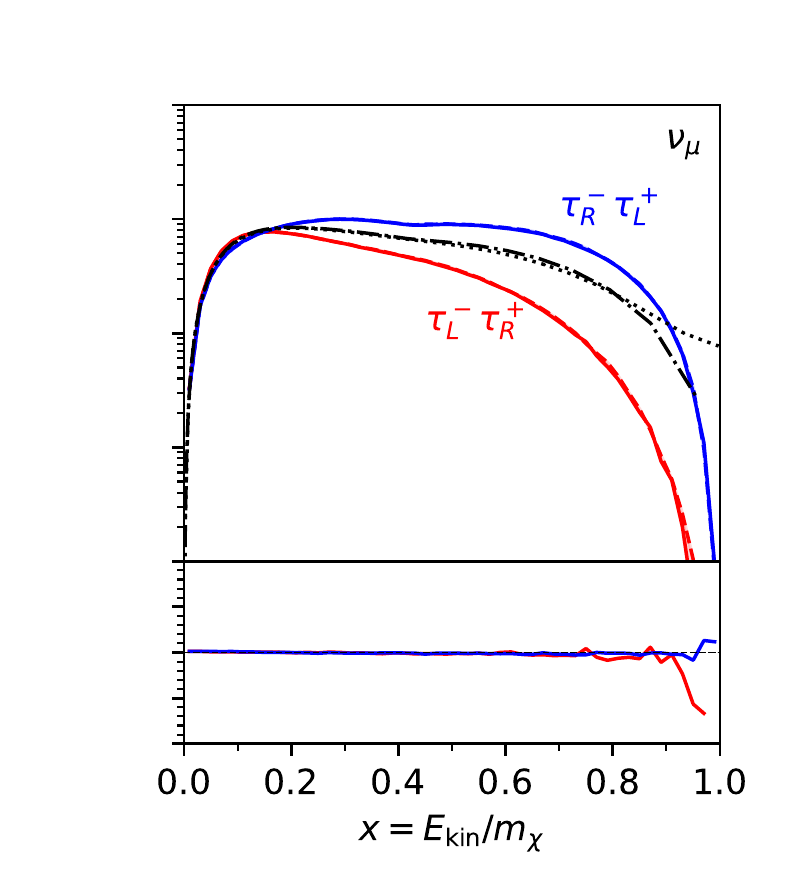}\\
    \includegraphics[height=0.3\textheight,clip,trim={0 0 0 8mm}]{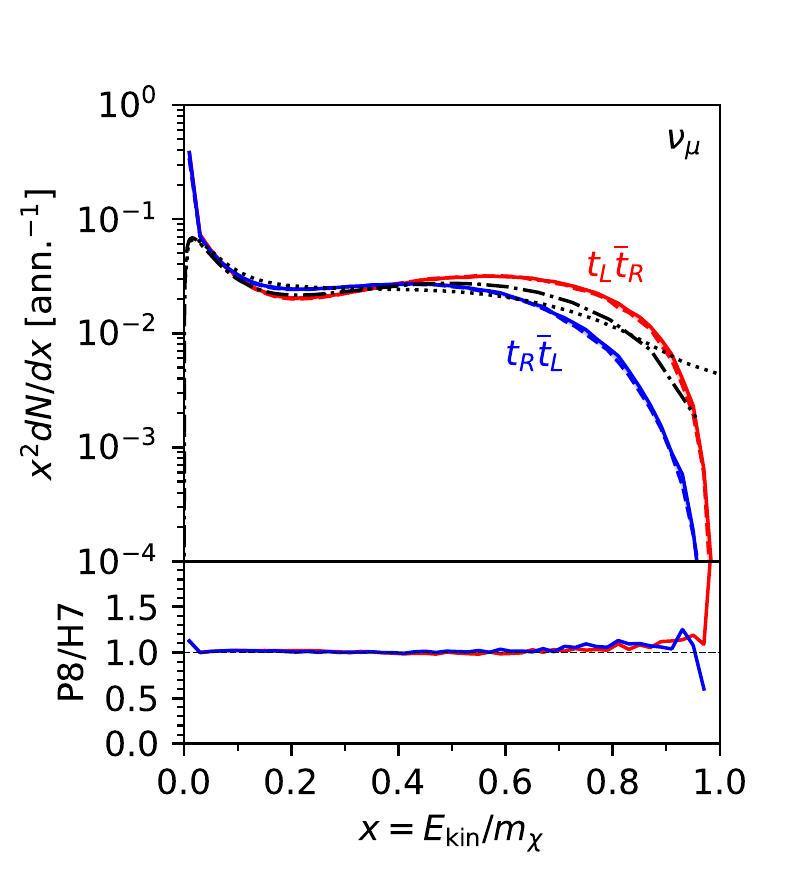}\hspace{-1.0em}
    \includegraphics[height=0.3\textheight,clip,trim={14mm 0 0 8mm}]{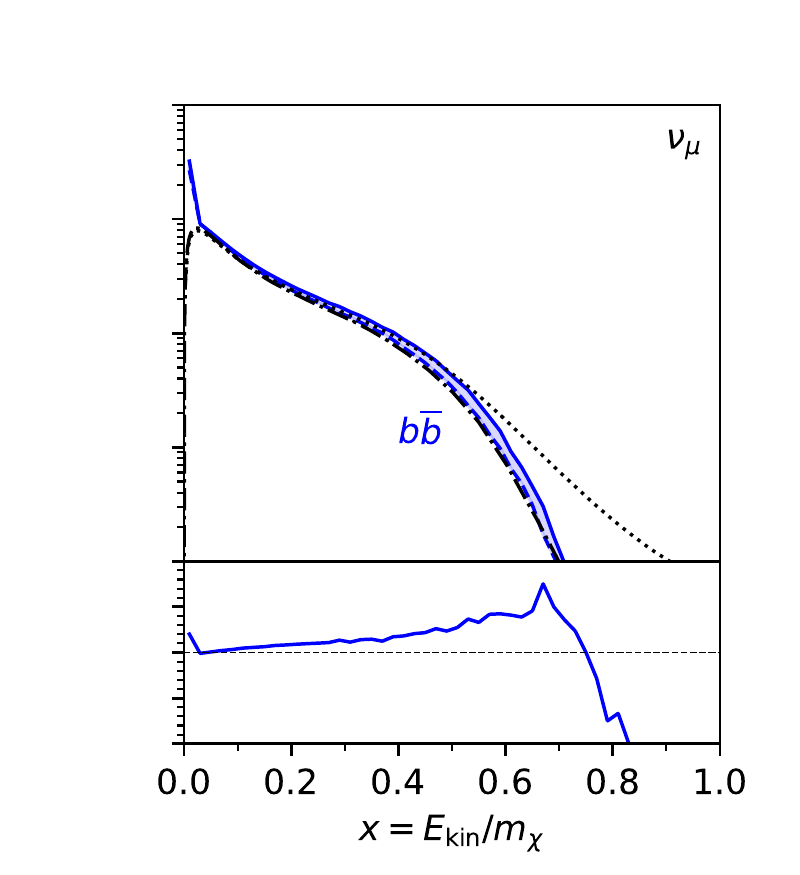}
    \caption{Yields of muon neutrinos (not including antineutrinos), comparing simulations done with \pythiae (solid) with simulations done with \herwigs (dashed). We also show the spectra for unpolarised final states simulated with \pythias (extracted from \darksusy) in the dash-dotted curves and the results of the \pppc \cite{Cirelli:2010xx} with dotted curves. The annihilation channels are indicated in each panel. The dark matter mass is here set to $m_\chi=\SI{1000}{\gev}$. In the plots, the upper part show $x^2dN/dx$ normalised to the number of simulated annihilations on the vertical axis whereas the lower part of each plot show the ratio, defined as $\text{\pe}/\text{\hs}$. The horisontal axis shows $x=E_{\rm kin}/m_\chi$ in all plots. }
    \label{fig:yields_nu_hvsp}
\end{figure}

\begin{figure}
    \centering
    \includegraphics[height=0.3\textheight,clip,trim={0 0 0 8mm}]{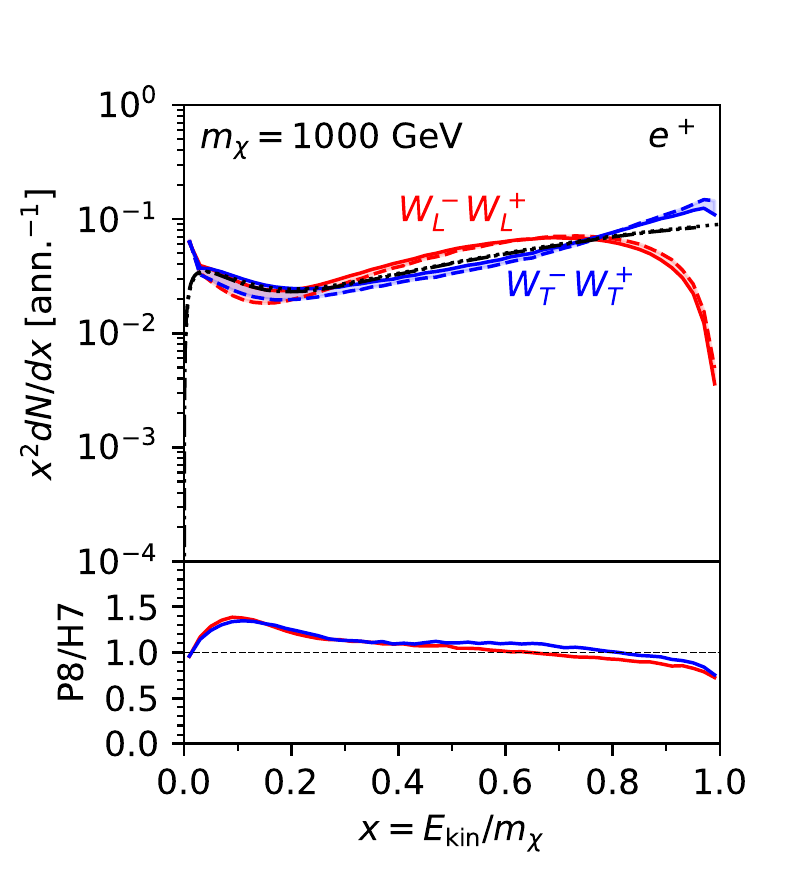}\hspace{-1.0em}
    \includegraphics[height=0.3\textheight,clip,trim={14mm 0 0 8mm}]{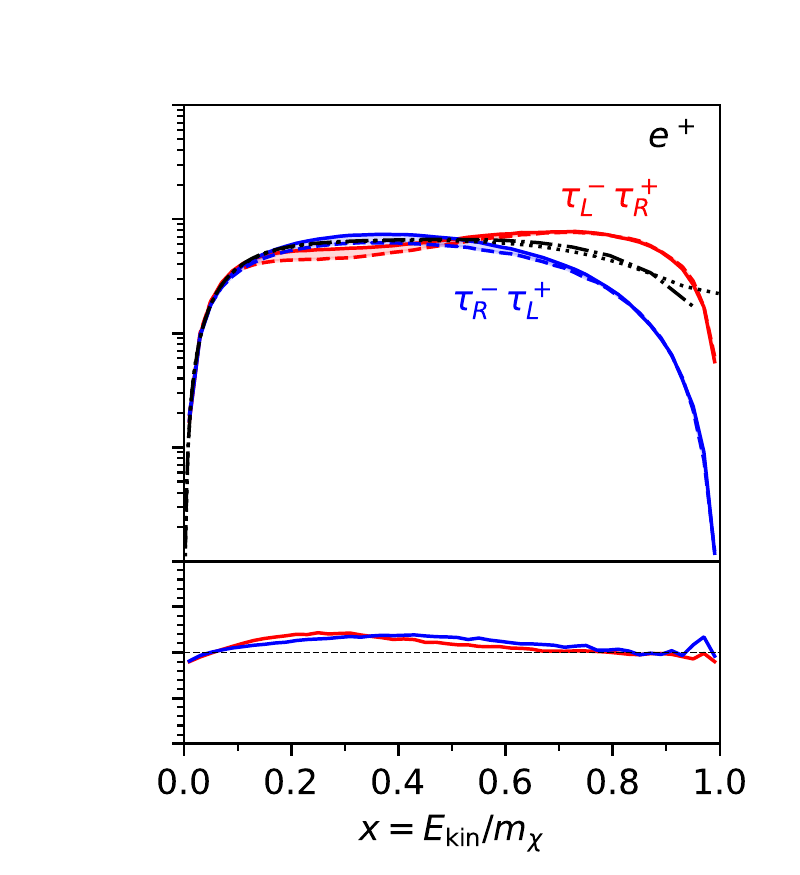} \\
    \includegraphics[height=0.3\textheight,clip,trim={0 0 0 8mm}]{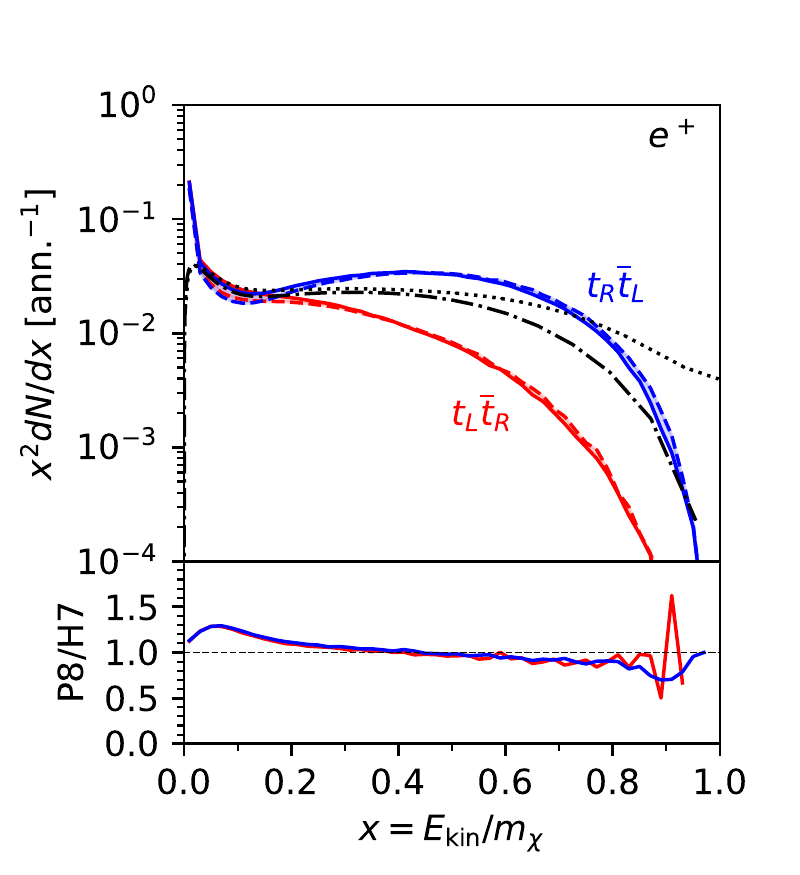}\hspace{-1.0em}
    \includegraphics[height=0.3\textheight,clip,trim={14mm 0 0 8mm}]{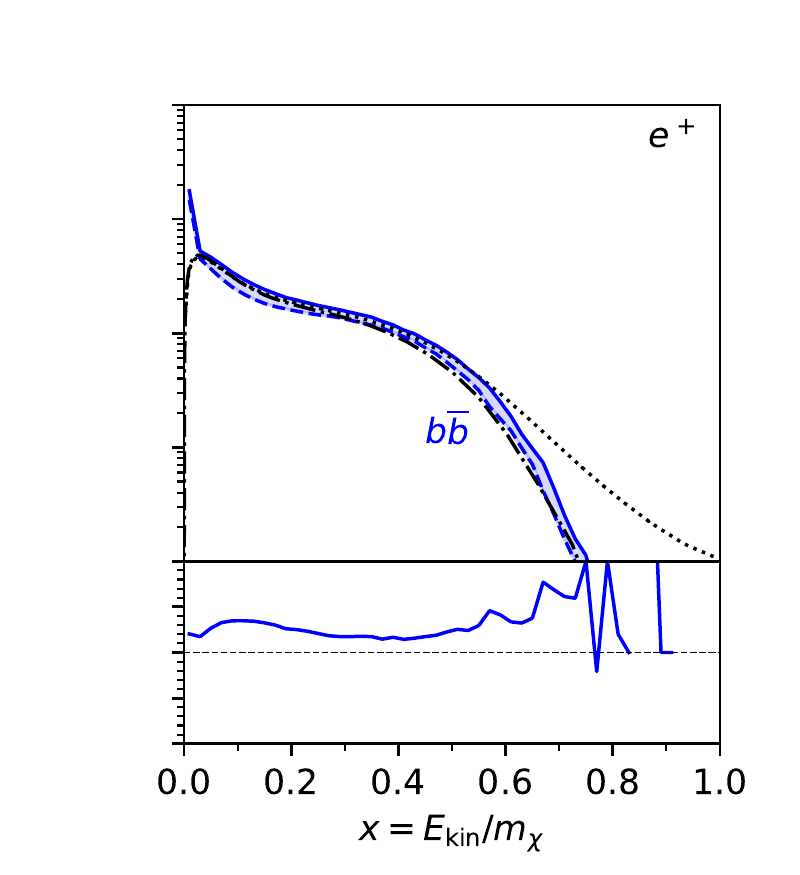}
    \caption{Same as figure~\ref{fig:yields_nu_hvsp} but for positron yields.}
    \label{fig:yields_ep_hvsp}
\end{figure}

\begin{figure}[t]
    \centering
    \includegraphics[height=0.3\textheight,clip,trim={0 0 0 8mm}]{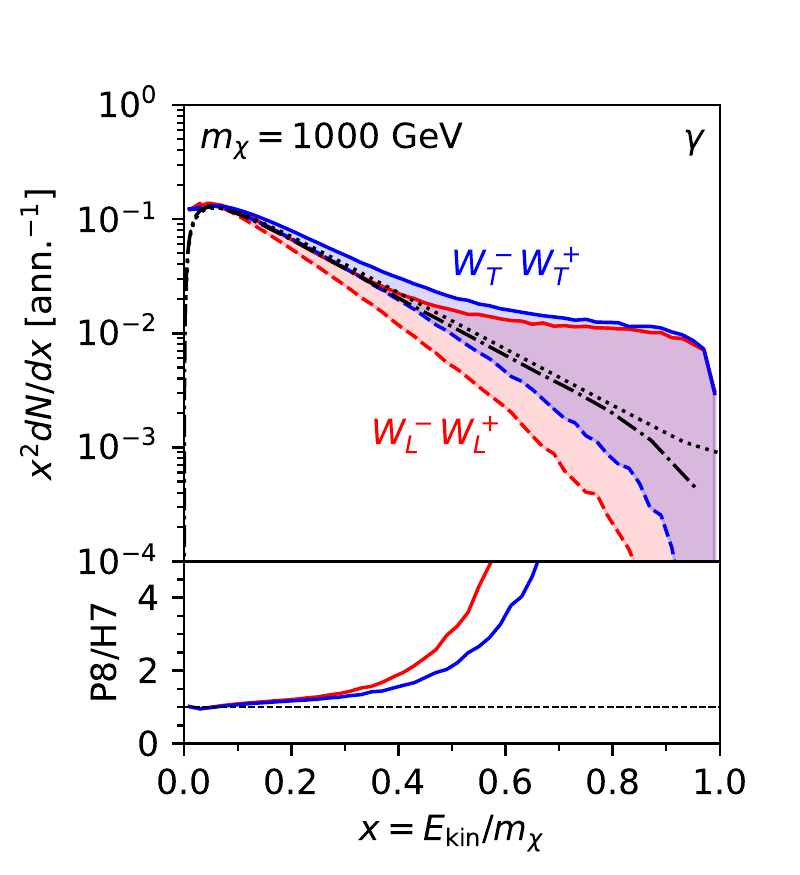}\hspace{-1.0em}
    \includegraphics[height=0.3\textheight,clip,trim={14mm 0 0 8mm}]{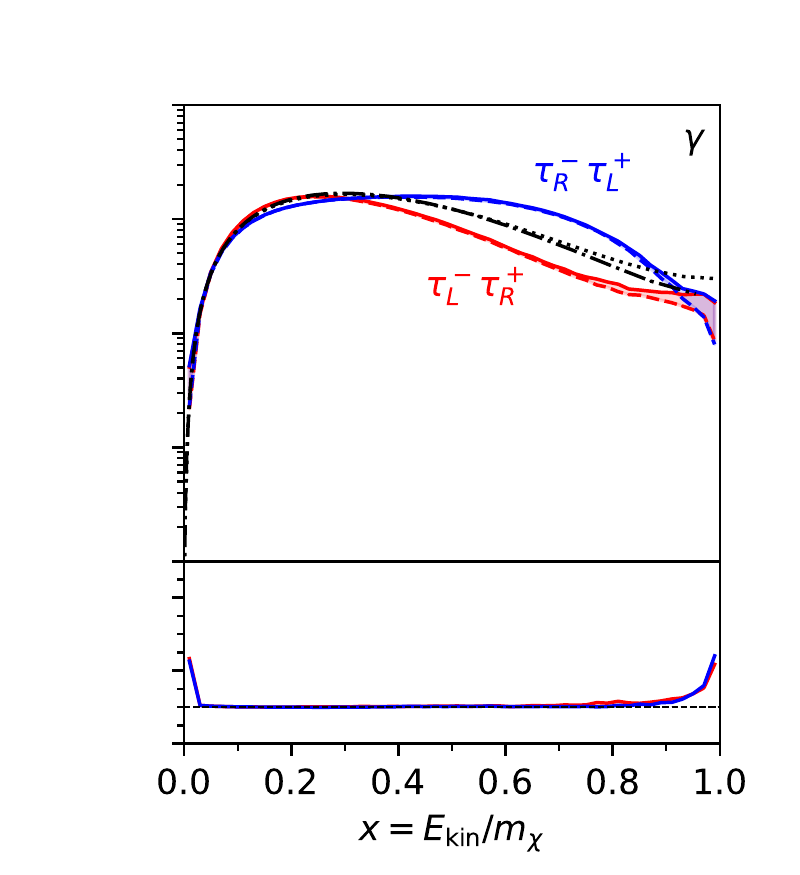}\\
    \includegraphics[height=0.3\textheight,clip,trim={0 0 0 8mm}]{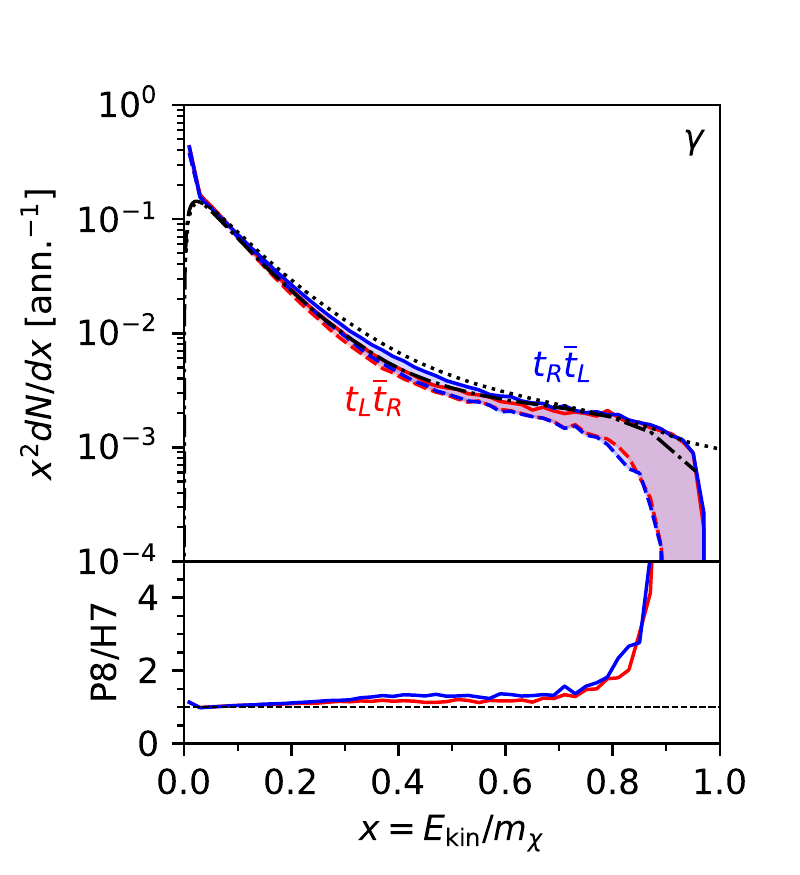}\hspace{-1.0em}
    \includegraphics[height=0.3\textheight,clip,trim={14mm 0 0 8mm}]{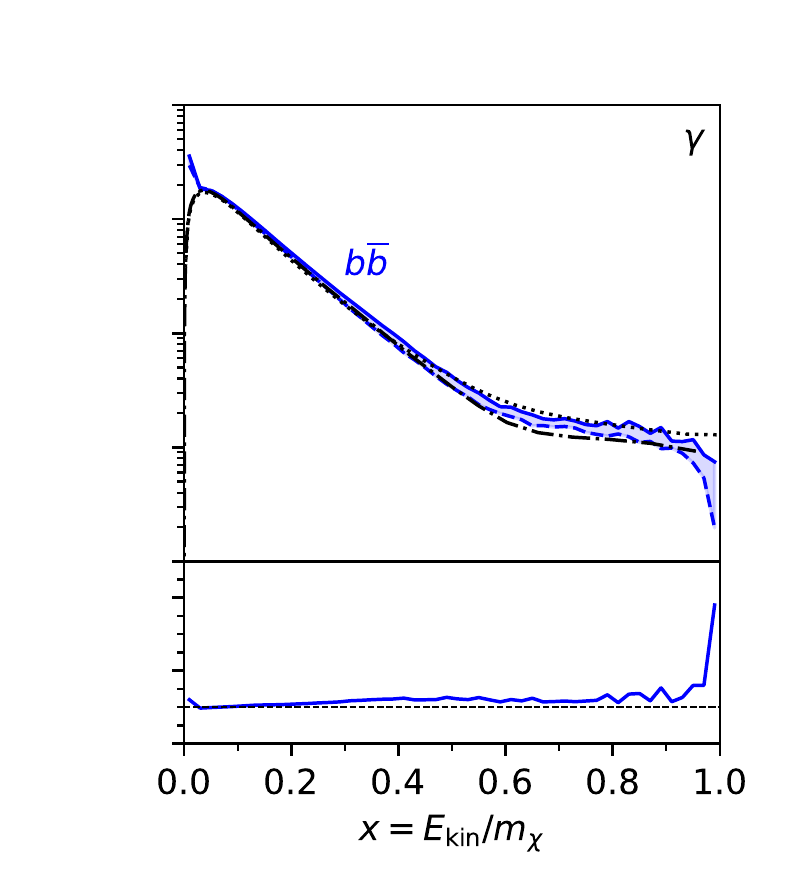}
    \caption{Same as figure~\ref{fig:yields_nu_hvsp} but for gamma-ray yields.}
    \label{fig:yields_gamma_hvsp}
\end{figure}

\begin{figure}
    \centering
    \includegraphics[height=0.3\textheight,clip,trim={0 0 0 8mm}]{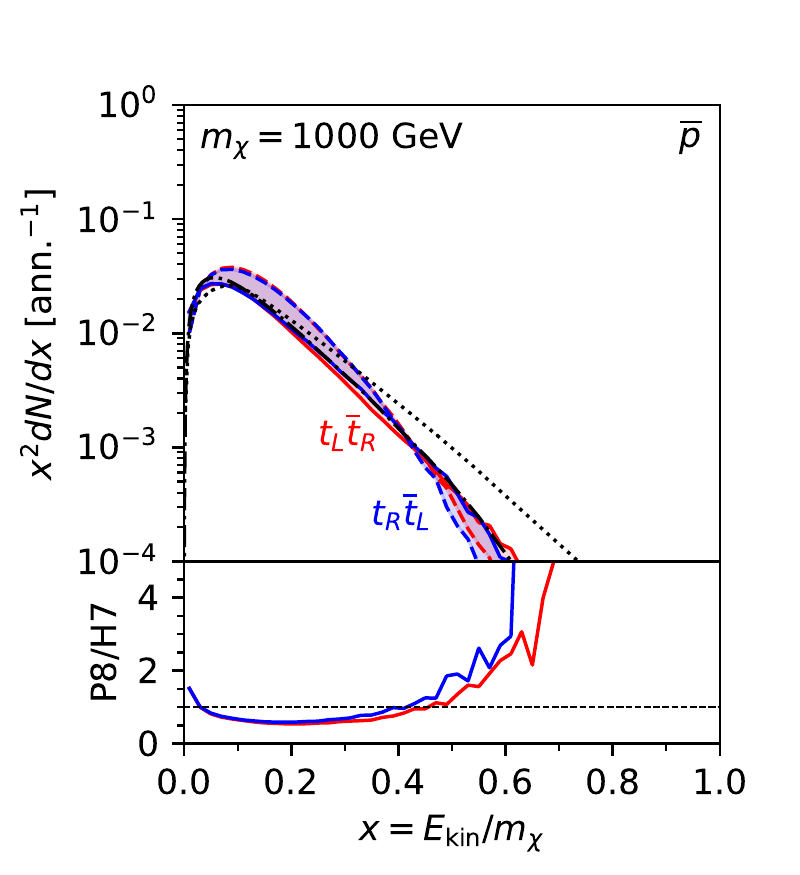}\hspace{-1.0em}
    \includegraphics[height=0.3\textheight,clip,trim={14mm 0 0 8mm}]{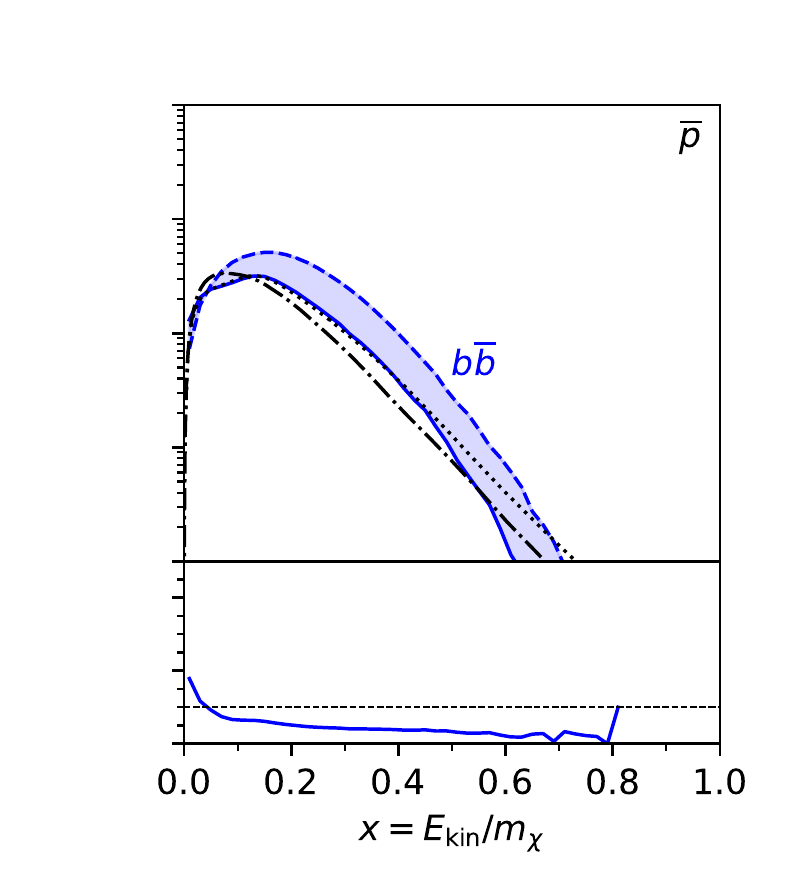}\\
    \includegraphics[height=0.3\textheight,clip,trim={0 0 0 8mm}]{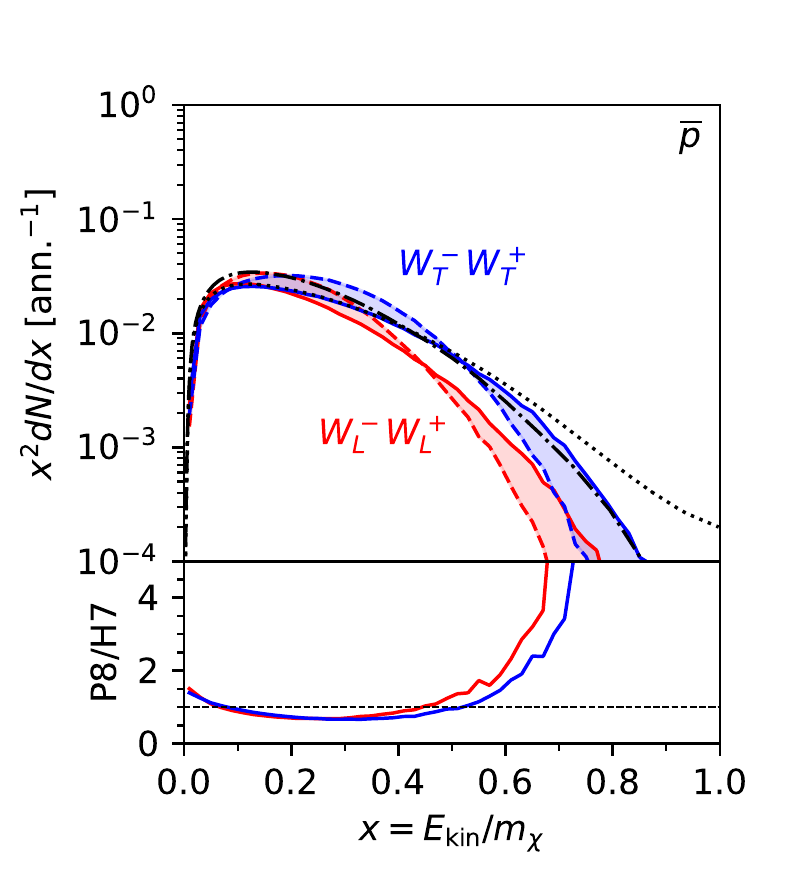}
    \caption{Same as figure~\ref{fig:yields_nu_hvsp} but for antiproton yields.}
    \label{fig:yields_pbar_hvsp}
\end{figure}

In figures~\ref{fig:yields_nu_hvsp},~\ref{fig:yields_ep_hvsp},~\ref{fig:yields_gamma_hvsp} and \ref{fig:yields_pbar_hvsp} we show the yields of muon neutrinos, positrons, gamma-rays and antiprotons for the simulated annihilation channels, here for a dark matter mass $m_\chi=\SI{1000}{\gev}$. We show results for simulations using \madgraph and \pythiae (referred to in the following as \pe) or \madgraph and \herwigs (\hs) and indicate the difference with the coloured band. As reference, we also show results extracted from \darksusy in dash-dotted lines, that have been simulated using \pythias (\ps) with unpolarised final states, and the results of ref.~\cite{Cirelli:2010xx} (\pppc) in dotted lines. 

In Appendix \ref{app:100GeV}, similar plots are shown for a dark matter mass of $m_\chi=\SI{100}{\gev}$. While the shape of the spectra are different at this lower energy, ultimately the ratios between the spectra produced are similar, with the exception of photons from the $W^+ W^-$ final state, as discussed in section~\ref{sec:gammas}. 


\subsubsection{Neutrino yields} 
We show in figure~\ref{fig:yields_nu_hvsp} the muon neutrino yields. For the neutrinos, the hardest spectra are produced in the gauge boson annihilation channels, since the gauge bosons can decay directly into leptons. In two-body decays such as $W^-\to \ell^- \overline{\nu}_\ell$, the leptons are monochromatic in the rest frame of the decaying gauge boson. When boosted to the annihilation rest frame, the spectrum is spread between energies $E_-$ and $E_+$, in the annihilation rest frame given by
\begin{equation}
E_{\pm}=\frac{m_\chi}{2}\left(1\pm\sqrt{1-\frac{m_W^2}{m_\chi^2}}\right).
\end{equation}
The shape of the spectrum between $E_-$ and $E_+$ depends on the polarisation of the decaying gauge boson, with the longitudinal (transverse) polarisation resulting in a convex (concave) shaped lepton spectrum. 

Annihilation into tau leptons and top quarks also produces relatively hard neutrino spectra, although somewhat softer than the gauge boson spectra. The softest neutrino spectra are produced in the $b\overline{b}$ channel, where the neutrinos predominantly come from decays of hadrons.

We note that in general, differences between \pe and \hs are small for neutrinos. Given that the main differences between the event generators lie in the handling of QCD effects, it is expected that there will be less differences in leptonic processes. As the \ps simulations are for unpolarised final states, their yields typically lie between those from polarised final states, as expected. The \pppc results show an excess at high energies relative to our results from both \pe and \hs, particularly in the quark channels.

\subsubsection{Positron yields} 
As one can see in figure~\ref{fig:yields_ep_hvsp}, the situation for positrons is similar as for the muon neutrinos. The gauge boson spectra are very similar at high energies, as expected, since these positrons also come mainly from direct $W$ boson decays. At lower energies there are some differences between \pe and \hs, with the \pe simulation giving a larger spectrum. This difference can most likely be traced to the differences in the hadronisation procedure, as the low energy positrons mainly come from decays of particles produced in the hadronisation step. 

As expected, the main differences between the event generators are found for the quark channels, primarily for the $b\overline{b}$ channel, where they reach around 20\%. Again the \pppc yields show an excess at high energies in the hadronic channels relative to our results.

\subsubsection{Gamma-ray yields}
\label{sec:gammas}
For the gamma-ray yields, shown in figure~\ref{fig:yields_gamma_hvsp}, there are large differences for the $W$ boson channels. The difference is related to the bremsstrahlung of photons from the $W$ bosons in the final state. The rate of bremsstrahlung in \hs is significantly lower than in \pe.\footnote{We have been unable to conclusively trace the source of this difference. We have seen in the simulations that in \herwigs, unlike in \pythiae, no photons at all seem to be emitted from the $W$ bosons.} We conclude that QED bremsstrahlung has a significant effect on the high energy part of the gamma-ray spectra. Reducing the energy of the $W^+ W^-$ pair of course reduces the amount of bremsstrahlung radiation, and so at lower dark matter masses, the agreement between the codes is better, as shown in figure~\ref{fig:yields_gamma_hvsp_100}.

For the other channels, differences are smaller, and mainly concentrated to the tails of the distributions. This is expected, since at high energies the event generators rely to a higher extent on extrapolation away from the tuning regions.

\subsubsection{Antiproton yields}
As expected, consistent differences between the generators are found in the antiproton yields, shown in figure~\ref{fig:yields_pbar_hvsp}. The ratio between the two are for most energies around a factor of two and the peaks of the distributions are slightly shifted, with \pe generally giving harder spectra than \hs, except in the $b\overline{b}$ channel, where the \hs spectrum is instead more energetic.

\subsection{Effect of polarisation}
\begin{figure}[t]
    \centering
    \includegraphics[height=0.3\textheight,clip,trim={0 0 0 8mm}]{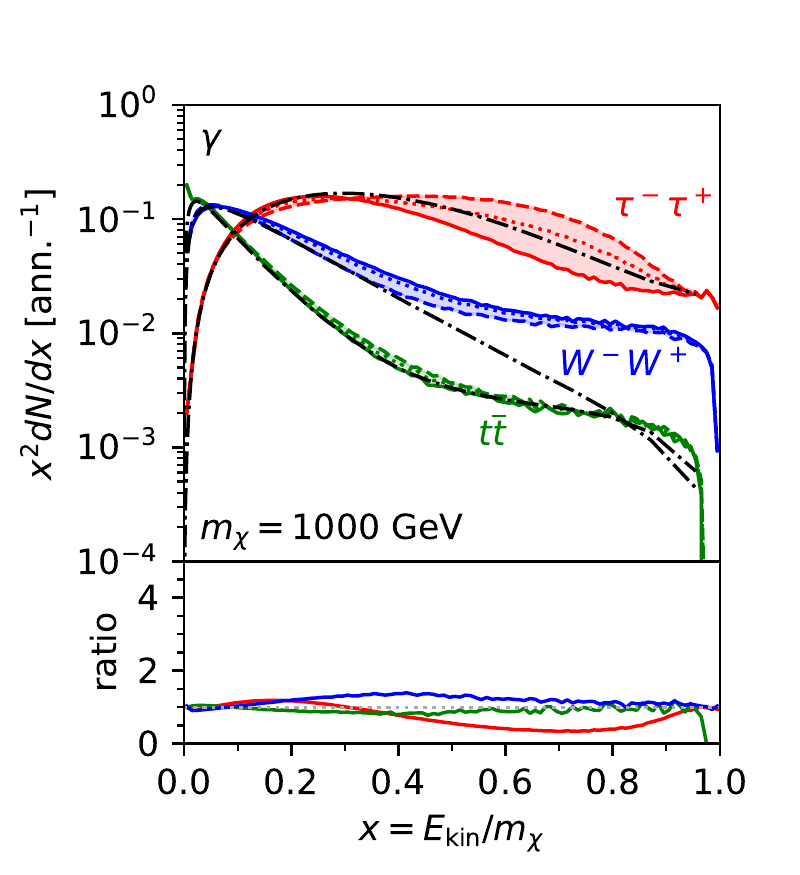}\hspace{-1.5em}
    \includegraphics[height=0.3\textheight,clip,trim={14mm 0 0 8mm}]{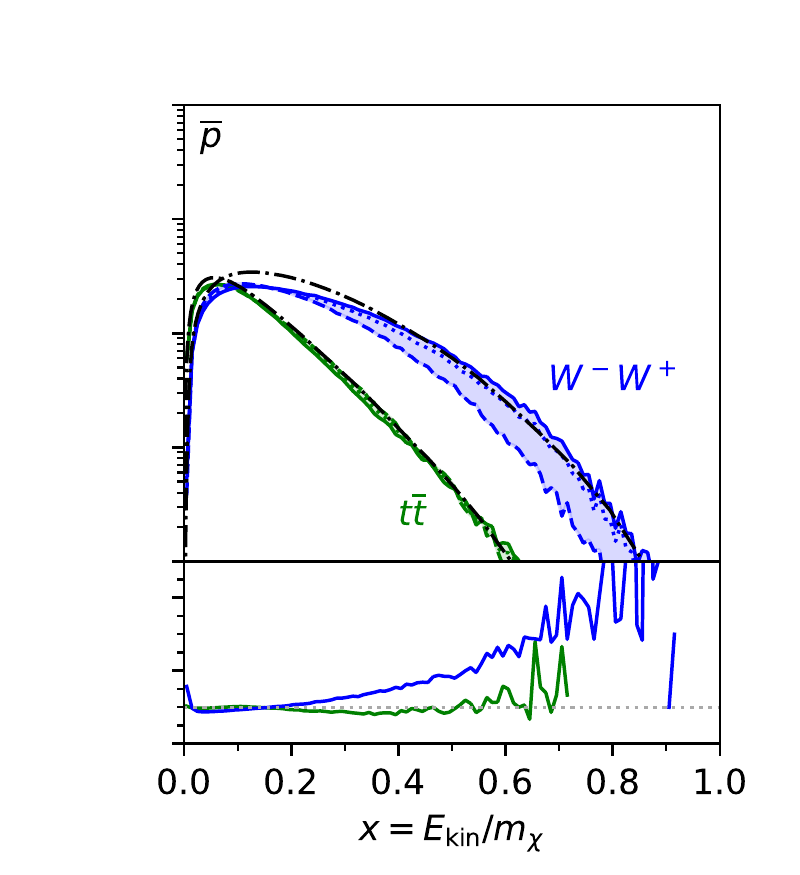}\\
    \includegraphics[height=0.3\textheight,clip,trim={0 0 0 8mm}]{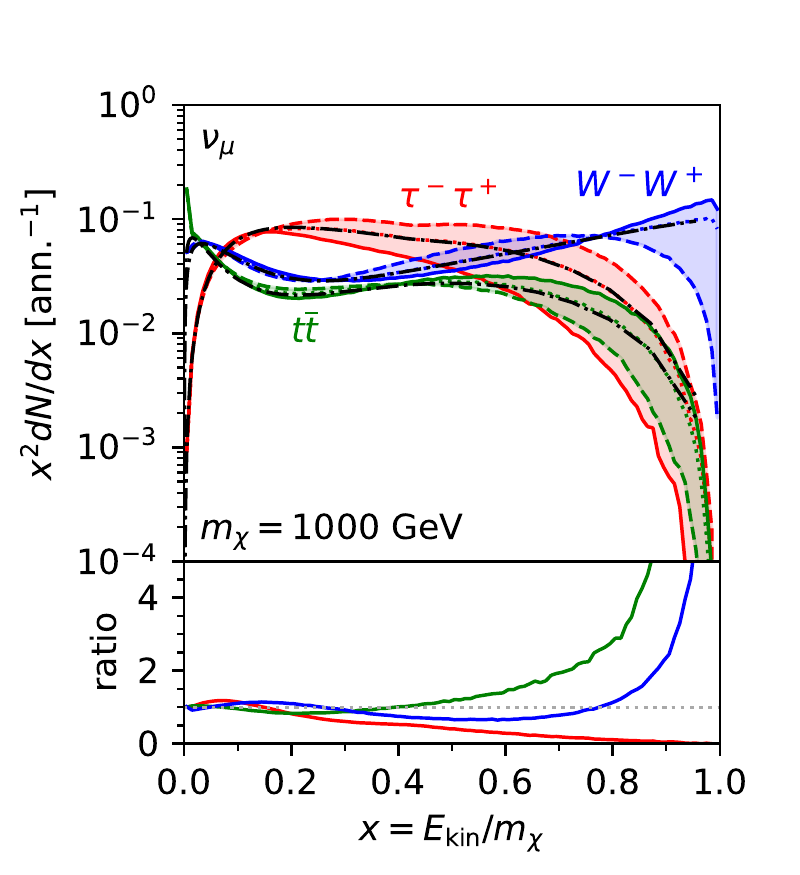}\hspace{-1.5em}
    \includegraphics[height=0.3\textheight,clip,trim={14mm 0 0 8mm}]{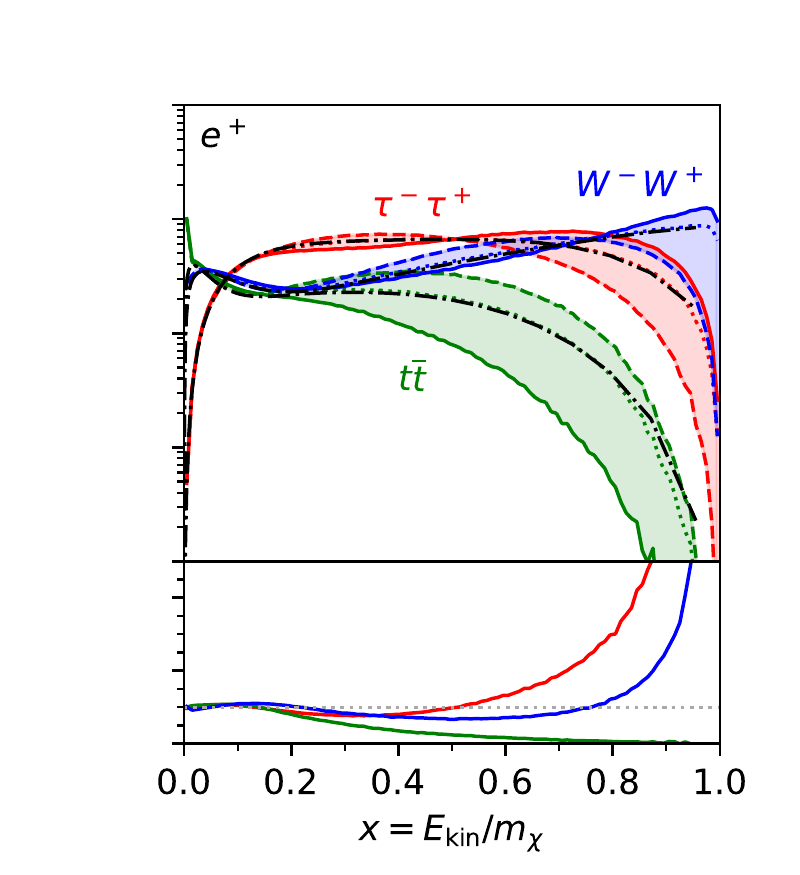}
    \caption{Yields of positrons, muon neutrinos, antiprotons and gamma-rays from dark matter annihilations, comparing simulations done with \pythiae for different polarisations of the same final state. We show also results using \pythias for unpolarised final states in the dash-dotted, black curves. The annihilation channels are indicated in each panel and the solid curves correspond to polarisation states in $\{\tau_L^- \tau_R^+, t_L \overline{t}_R, W_T^- W_T^+\}$ whereas the dashed curves correspond to states in $\{\tau_R^- \tau_L^+, t_R \overline{t}_L, W_L^- W_L^+\}$ and the dotted curves are the unpolarised average of these. The dark matter mass is here set to $m_\chi=\SI{1000}{\gev}$. In the plots, the upper part show $x^2dN/dx$ normalised to the number of simulated annihilations whereas the lower part of each plot show the ratio, defined as $\text{solid}/\text{dashed}$. The horizontal axis shows $x=E_{\rm kin}/m_\chi$ in all plots.  } 
    \label{fig:pyt-pol-m1000}
\end{figure}

 We show in figure~\ref{fig:pyt-pol-m1000} yields of positrons, muon neutrinos, antiprotons and gamma-rays for different polarisations in the final state, with the difference between the polarisations highlighted by the coloured band, and the average represented by the dotted line of the same colour. An unpolarised final state will follow this line in the centre of the band. We show here results obtained with the \pe simulation, compared with simulations performed with \ps (for unpolarised final states) in the dash-dotted curves. The coloured band in this case represents the effect that the polarisation of the final states can have on the yields. 
 
We see in figure~\ref{fig:pyt-pol-m1000} that the differences between different polarisations of the final states are most apparent at higher energies. This is expected, since the polarisation in our simulations affects the decays occurring earliest in the decay chain. For example, we can clearly see the expected concave and convex shape of the neutrino spectrum from $W_T^- W_T^+ $ and $W_L^- W_L^+$ final states respectively, with the spectrum in the $W_T^- W_T^+$ final state clearly peaking at the highest energies. We also see, as expected, the unpolarised \ps result following closely the average between the two polarisations apart from the gamma-ray spectrum in the  $WW$ channel. The difference between \pythias and \pythiae is here most probably linked to a lack of QED bremsstrahlung in \pythias.

We can also see differences in the gamma-ray and antiproton spectra. The differences for gamma-rays can likely be attributed to the different energy spectra of the charged particles which radiate the photons. We can also see that, in particular for the antiproton spectra, there are differences between the polarisation average and the \ps result. This likely depends on the fact that the hadronisation and showering has changed between \ps and \pe.

\section{Impact on dark matter searches with neutrino telescopes}
For the neutrino fluxes, polarisation in the final state can have an important impact on dark matter searches. The polarisation of the final state results in differences in the high-energy part of the spectra, as shown in figure~\ref{fig:pyt-pol-m1000} above. This difference is further enhanced for the case of neutrino searches since the detectability of neutrinos in a neutrino telescope scales roughly as the flux times the neutrino energy squared, $E_\nu^2$, where one factor of $E_\nu$ comes from the rise of the neutrino interaction cross section with $E_\nu$ and the other from the muon range in the detector, which rises approximately linearly with muon energy, and hence is proportional to $E_\nu$. In addition, the main background in dark matter searches, the atmospheric neutrino flux, decreases sharply with energy. Therefore, the high end part of the flux becomes highly relevant for dark matter searches with neutrino telescopes. For this reason we focus in this section on the impact of polarisation on the neutrino fluxes specifically. 

\subsection{Simulation of neutrino event rates with \textsf{WimpSim}} 
To make predictions of event rates in neutrino telescopes for different polarisations we use the \wimpsim package \cite{Blennow:2007tw,Edsjo:2017kjk,Niblaeus:2019gjk}.  \wimpsim is an event based \textsf{Fortran} code that simulates the full chain from dark matter annihilation to the interaction in a neutrino telescope on Earth. The neutrinos are generated in dark matter annihilations in the centre of the Sun or Earth. The produced neutrinos are then propagated from the production point to a detector on Earth, where interactions in the Sun or Earth and oscillations are handled in a full three-flavour setup. Apart from simulating dark matter annihilations in the centre of the Sun or Earth \cite{Blennow:2007tw}, there are also options to simulate the neutrino flux from cosmic ray interactions in the solar atmosphere \cite{Edsjo:2017kjk} as well as the flux of neutrinos, gamma-rays and charged cosmic rays from annihilation of dark matter into long-lived mediators that decay away from the solar core \cite{Niblaeus:2019gjk}. 

Traditionally, \wimpsim has used \pythias for the simulation of the dark matter annihilations. For this study, we primarily want to study other event generators (namely \pythiae and \herwigs) and have therefore added a new functionality in \wimpsim to make it possible to read in event files with neutrino yields at production. This allows us to study the effect the different spectra from our simulations will have on the event rates in neutrino telescopes. We have chosen to do this part of the analysis using only the \pythiae simulation data, but results using \herwigs will be similar given that the differences between the generators is generally small for neutrino spectra. For annihilations in the Sun, we have taken care not to include neutrinos coming from the decays of muons, neutrons or charged kaons and pions, as these particles are quickly stopped in the solar interior and do not give rise to high energy neutrinos.

For neutrinos coming from heavy hadrons containing $b$ or $c$-quarks, one should take into account the interactions of the hadrons with the solar material and reduce the neutrino energies correspondingly. Currently we have not included this when reading event files into \wimpsim for the production fluxes. This effect will be most important for final states giving many of these heavy hadrons (for example the $b\overline{b}$ channel) and since we focus on $W$ boson and $\tau$ lepton channels for this part of the study, it will not have a large effect. 

The annihilation point is sampled from a thermal distribution around the solar core. Assuming a thermal annihilation distribution has been shown to be a very good approximation \cite{Widmark:2017yvd}. The neutrinos are then allowed to interact through charged current (CC) or neutral current (NC) interactions and oscillate (including matter effects) on their way from the annihilation point to the detector. If a $\tau^\pm$ is produced in a CC interaction, the $\tau^\pm$ is decayed and any neutrinos from the decay are collected and included in the remaining neutrino propagation. For this decay, \pythias is used. As we are here interested in polarised final states in the dark matter annihilation, the fact that the \pythias $\tau$ decays do not include polarisation effects is not relevant for us.

As soon as $m_\chi$ is large enough it is essential to take into account the CC and NC interactions when simulating the neutrino fluxes. In the CC interactions, neutrinos are effectively lost, leading to an attenuation of the flux at high energies (since the cross section rises with the neutrino energy). The main effect of the NC interactions is to shift the neutrino flux to lower energies.

In the case of annihilations in the Sun, neutrinos are propagated to a distance of \SI{1}{\au} from the centre of the Sun. They are then propagated from \SI{1}{\au} to a specified detector location on Earth, taking into account the eccentricity of the Earth's orbit, and allowed to interact in the detector volume to produce a neutrino signal in the telescope. 

For neutrino oscillations and interactions we need values for the neutrino oscillation parameters, and furthermore need a solar model. We have used the best-fit values of the neutrino oscillation parameters from the \textsf{NuFit-3.2} global fit \cite{Esteban:2016qun, nufitonline} (summarised in table~\ref{tab:oscparams}). We consider only the normal mass hierarchy here. For the solar model, we use the model by Serenelli \textit{et al.} \cite{Serenelli:2009yc} which is the default in \darksusy.

\renewcommand{\arraystretch}{1.2}
\begin{table}[t]
    \centering
    \begin{tabular}{cc}
        \toprule \toprule
         Parameter & Best-fit value  \\ \midrule
         $\theta_{12}$ & \SI{33.62}{\degree} \\
         $\theta_{23}$ & \SI{47.2}{\degree} \\
         $\theta_{13}$ & \SI{8.54}{\degree} \\
         $\delta_{\rm{CP}}$ & \SI{234}{\degree} \\
         $\Delta m_{21}^2$ & \SI{7.40e-5}{\electronvolt\squared}   \\
         $\Delta m_{31}^2 $ & \SI{2.494e-3}{\electronvolt\squared} \\ \bottomrule \bottomrule
    \end{tabular}
    \caption{The values of the neutrinos oscillation parameters that we use in our \wimpsim simulations. These are the best-fit values for the normal mass hierarchy from the  \textsf{NuFit-3.2} global fit \cite{Esteban:2016qun,nufitonline}.}
    \label{tab:oscparams}
\end{table}
\renewcommand{\arraystretch}{1}

\section{Neutrino telescope results}
\begin{figure}[tp]
    \centering
    \includegraphics[height=0.3\textheight]{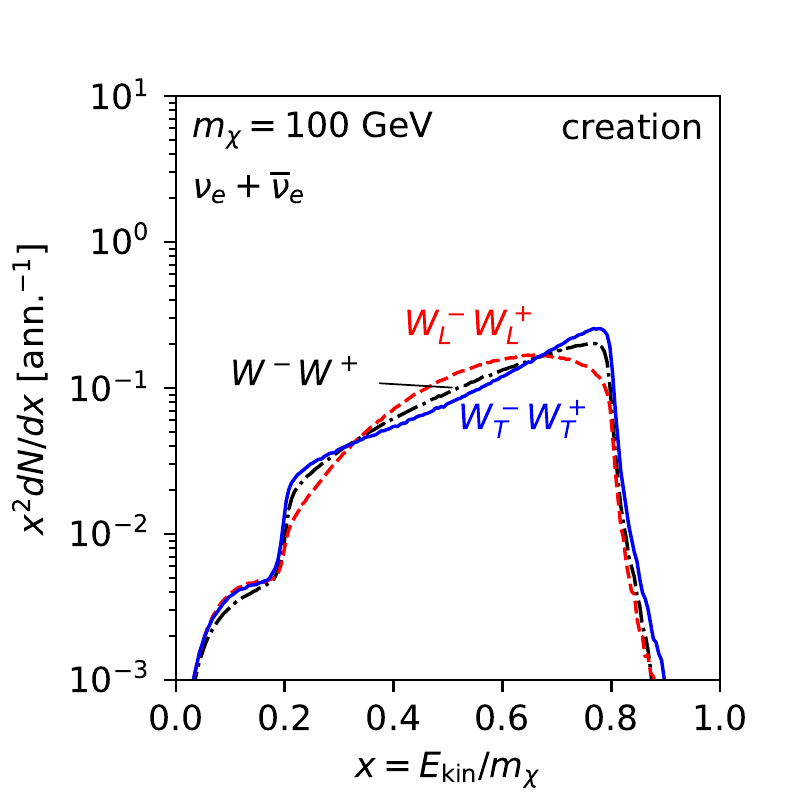}
    \includegraphics[height=0.3\textheight,clip,trim={7mm 0 0 0}]{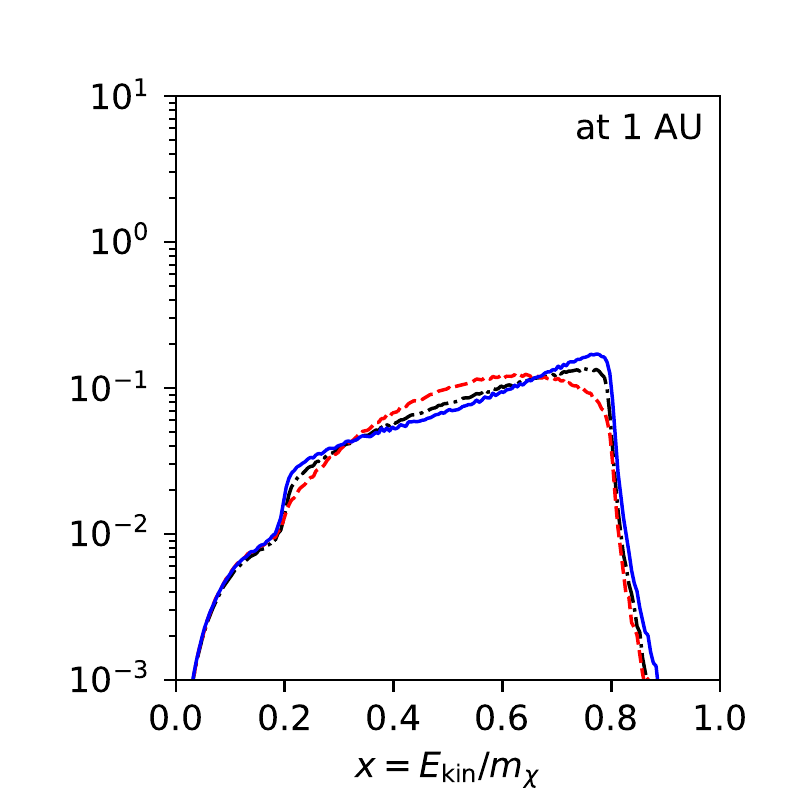} \\
    \includegraphics[height=0.3\textheight]{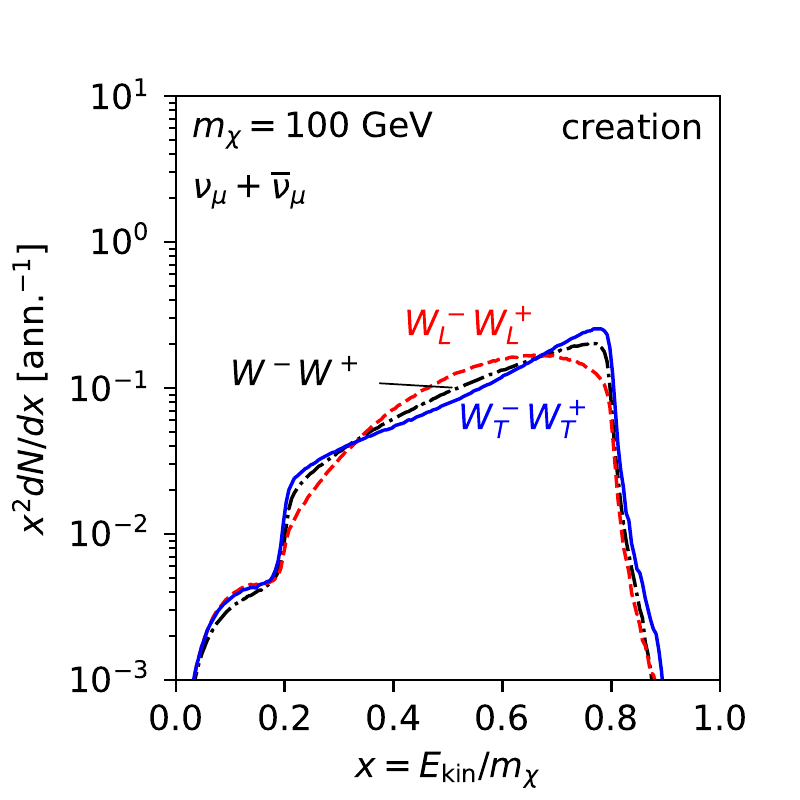}
    \includegraphics[height=0.3\textheight,clip,trim={7mm 0 0 0}]{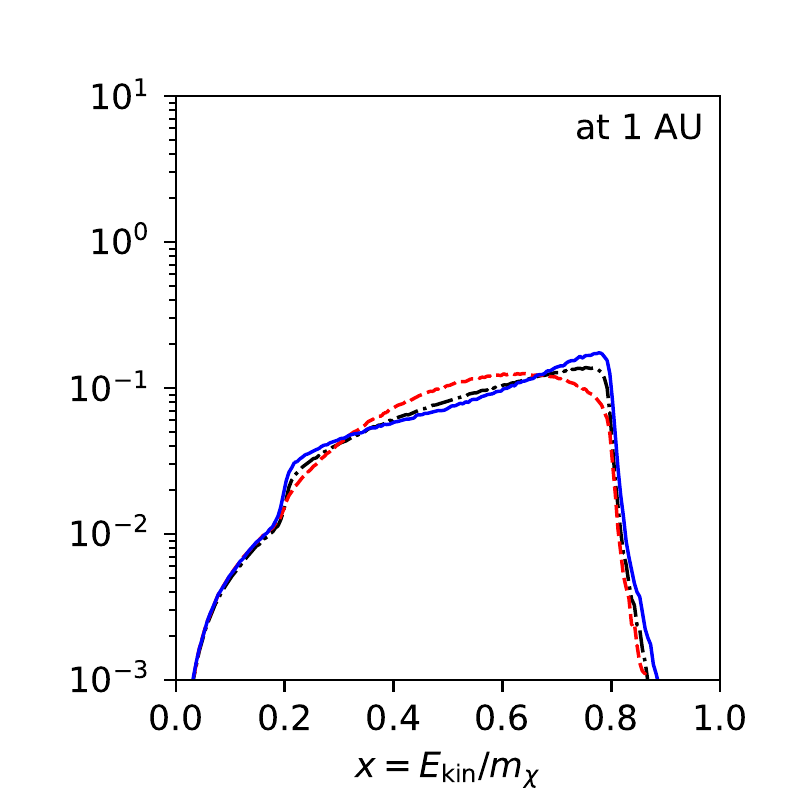} \\
    \includegraphics[height=0.3\textheight]{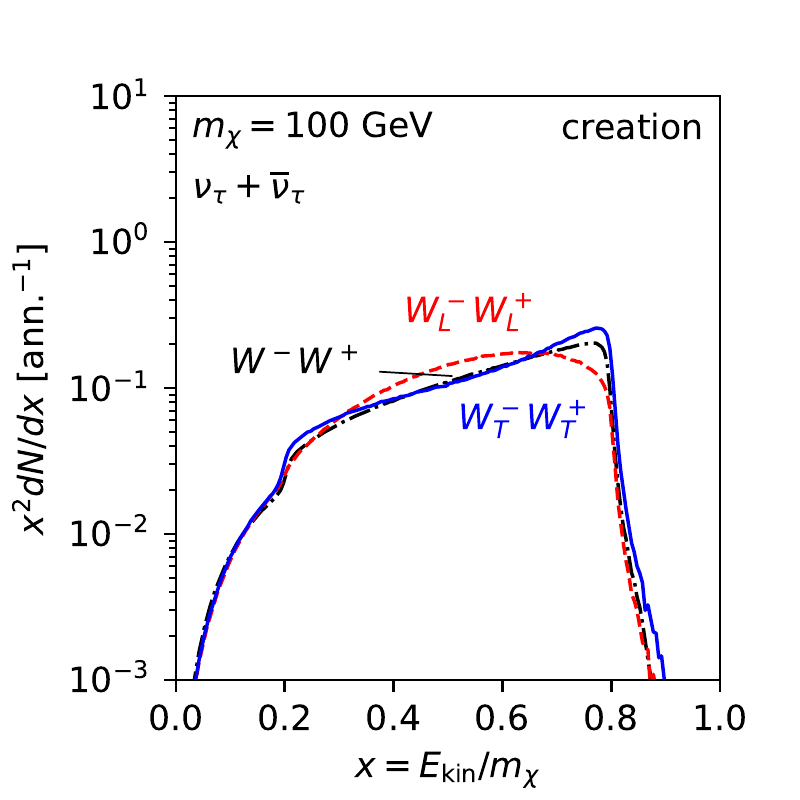}
    \includegraphics[height=0.3\textheight,clip,trim={7mm 0 0 0}]{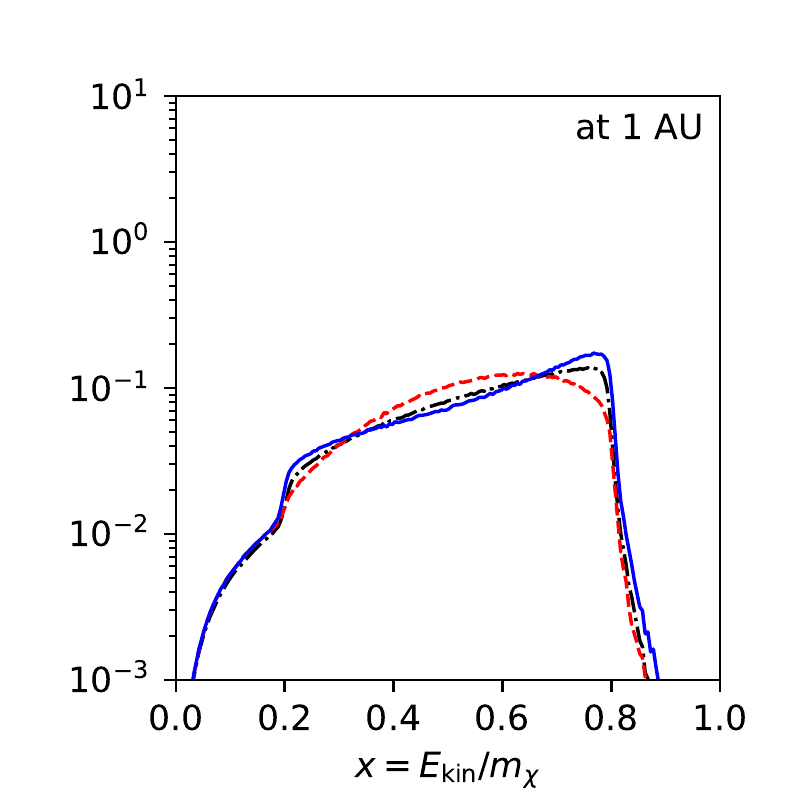}
    \caption{Sum of neutrino and antineutrino fluxes at production (left) and at \SI{1}{\au} from the Sun (right) for the $W$ boson annihilation channels with a dark matter mass of \SI{100}{\gev}. From top to bottom, the plots show the $\nu_e+\overline{\nu}_e$, $\nu_\mu+\overline{\nu}_\mu$ and $\nu_\tau+\overline{\nu}_\tau$ flux and we show in each plot the $W_T^- W_T^+$ channel, the $W_L^- W_L^+$ channel and the unpolarised $WW$ channel. The polarised channels use \pe results and the unpolarised channel is simulated with \pythias as per \wimpsim default running. }
    \label{fig:ww_m100}
\end{figure}

\begin{figure}[tp]
    \centering
    \includegraphics[height=0.3\textheight]{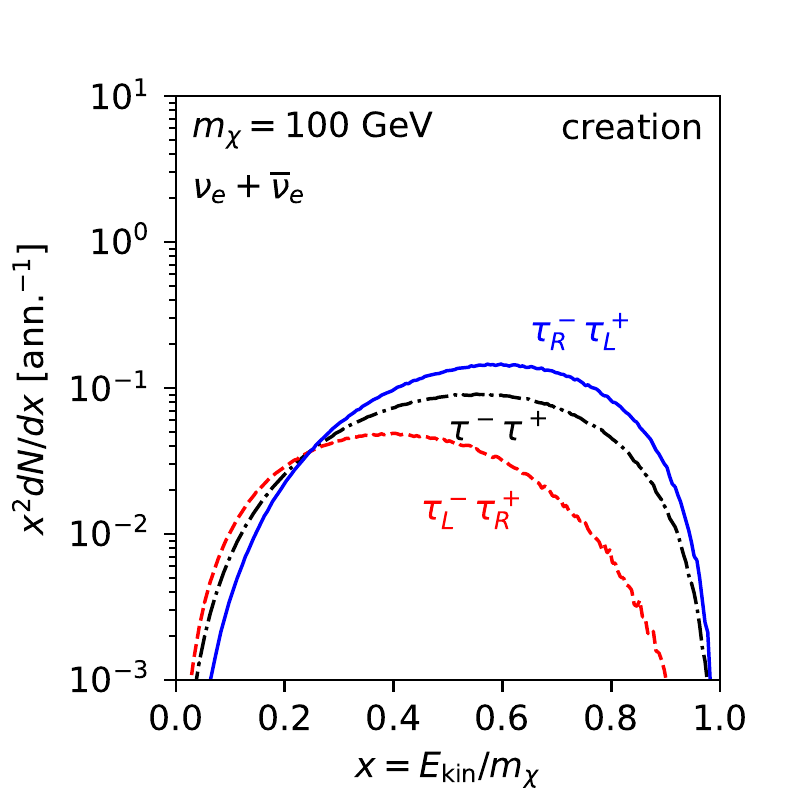}
    \includegraphics[height=0.3\textheight,clip,trim={7mm 0 0 0}]{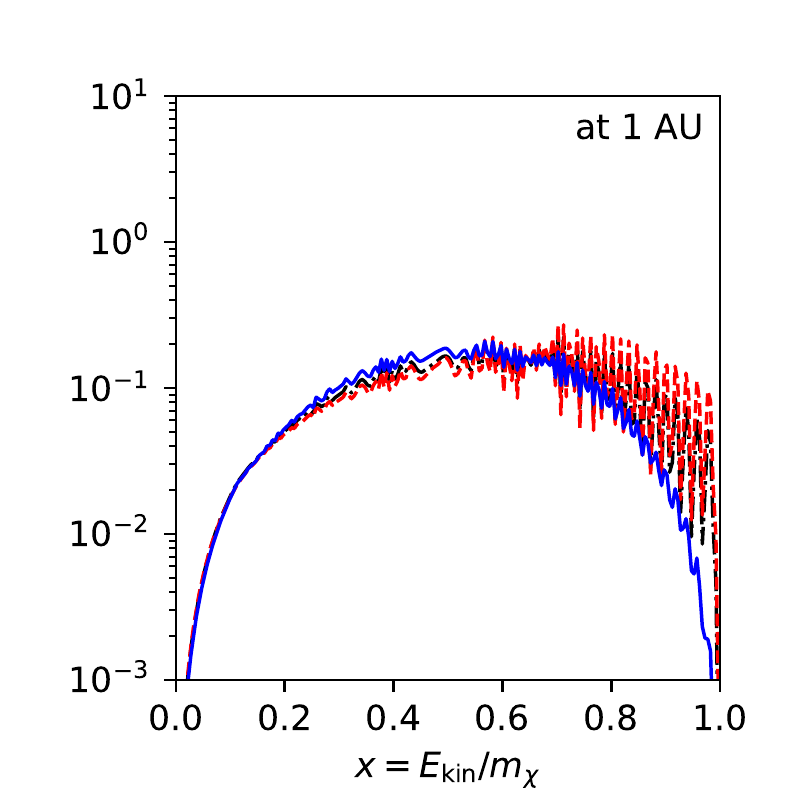} \\
    \includegraphics[height=0.3\textheight]{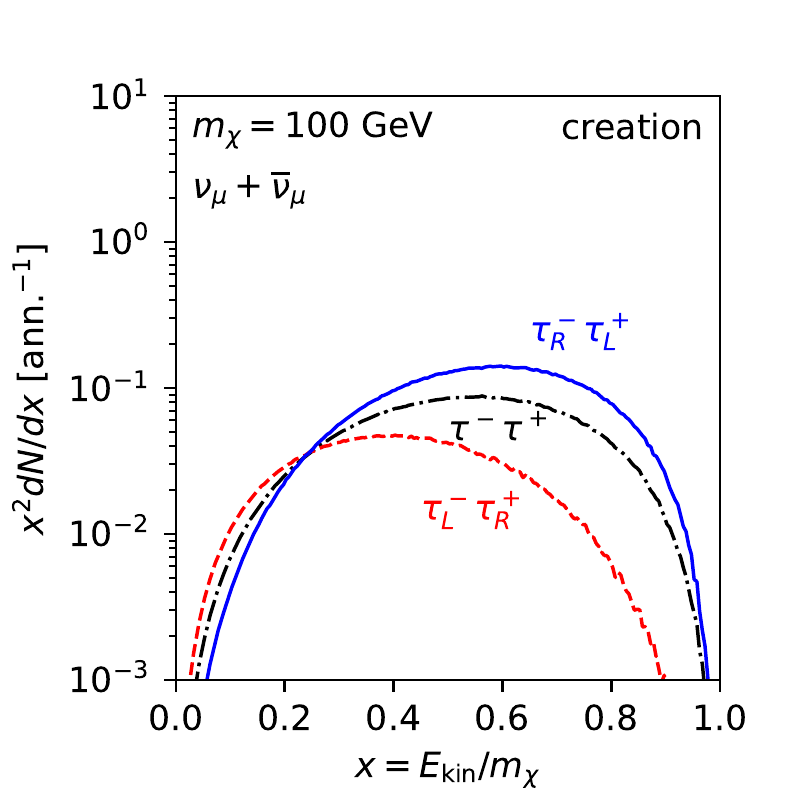}
    \includegraphics[height=0.3\textheight,clip,trim={7mm 0 0 0}]{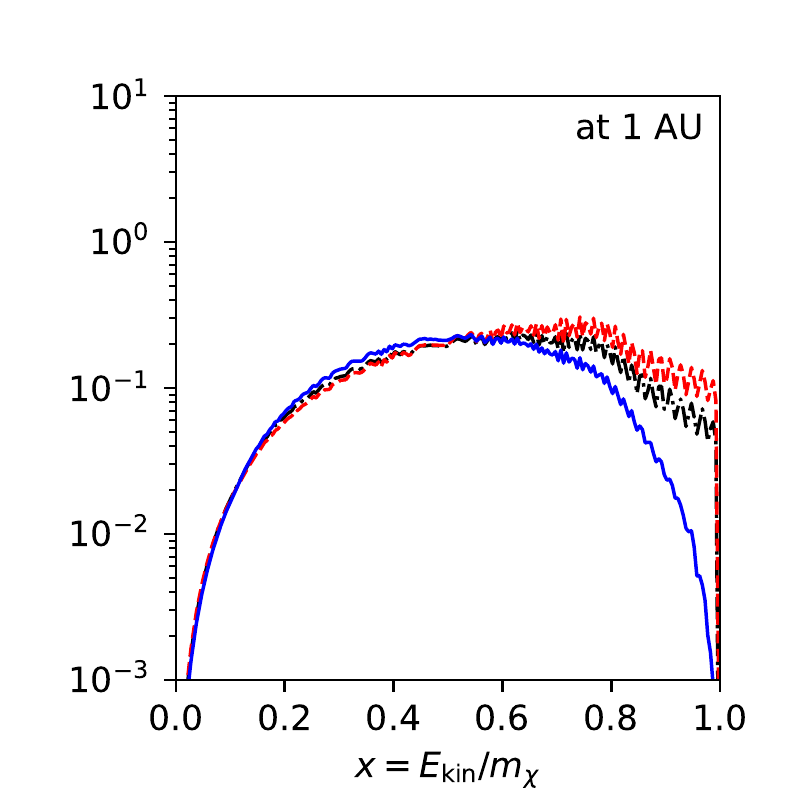} \\
    \includegraphics[height=0.3\textheight]{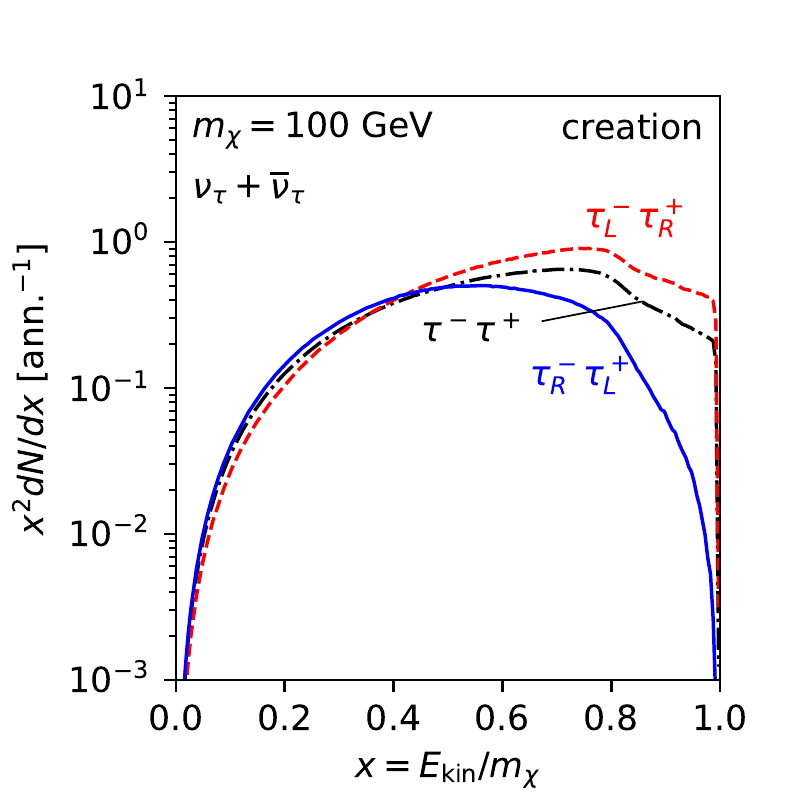}
    \includegraphics[height=0.3\textheight,clip,trim={7mm 0 0 0}]{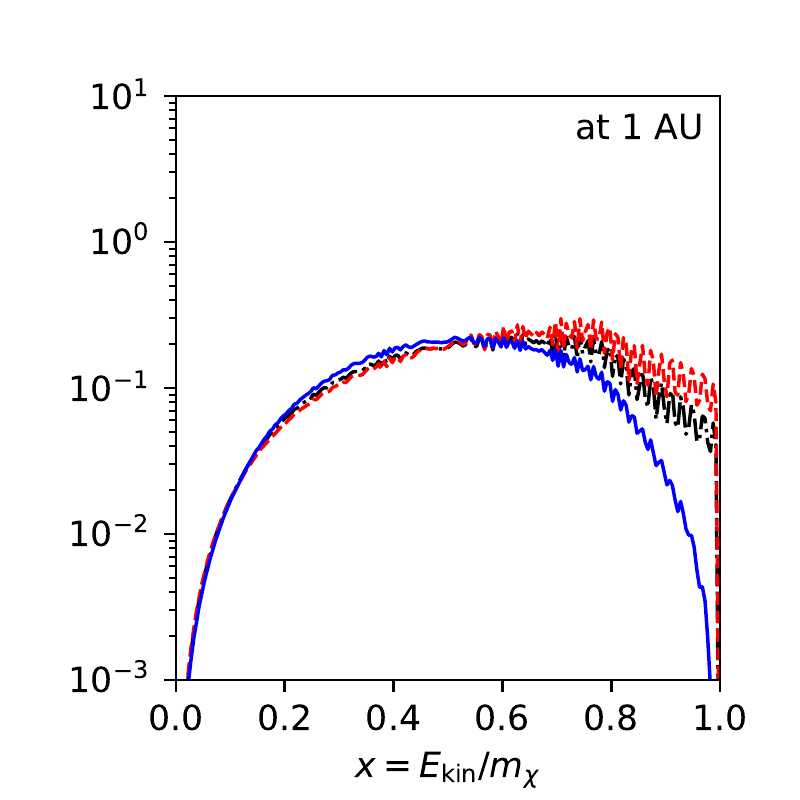}
    \caption{Sum of neutrino and antineutrino fluxes at production (left) and at \SI{1}{\au} from the Sun (right) for the $\tau$ annihilation channels with a dark matter mass of \SI{100}{\gev}. The curve belonging to each channel is indicated in the left plot (the same colour and line type is used in the right plots). From top to bottom, the plots show the $\nu_e+\overline{\nu}_e$, $\nu_\mu+\overline{\nu}_\mu$ and $\nu_\tau+\overline{\nu}_\tau$ flux and we show in each plot the $\tau_L^- \tau_R^+$ channel, the $\tau_R^- \tau_L^+$ channel and the unpolarised $\tau^- \tau^+$ channel. The polarised channels use \pe results and the unpolarised channel is simulated with \pythias as per \wimpsim default running. }
    \label{fig:tau_m100}
\end{figure}

\begin{figure}[tp]
    \centering
    \includegraphics[height=0.3\textheight]{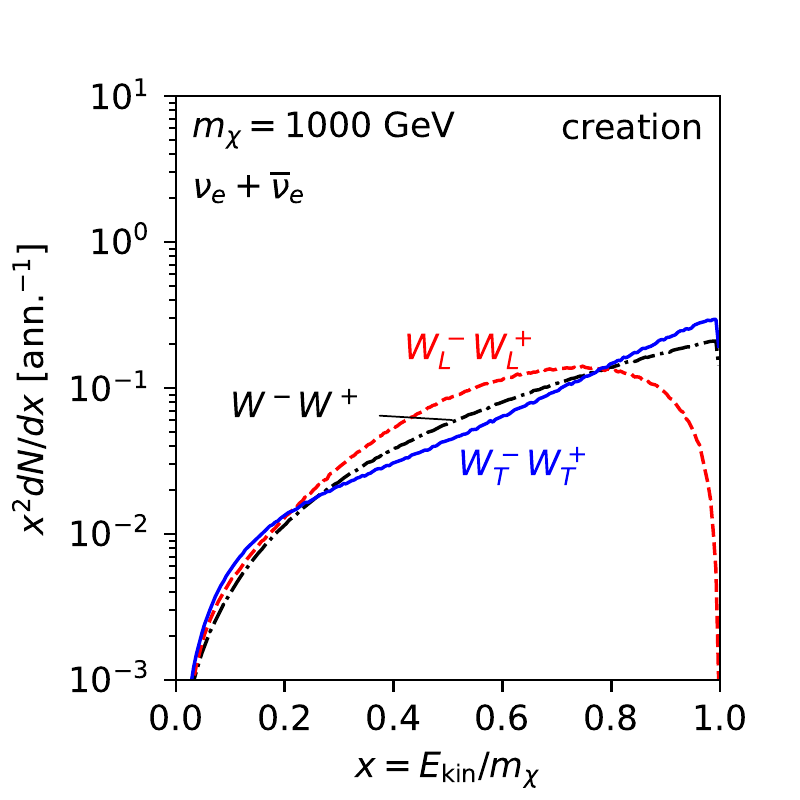}
    \includegraphics[height=0.3\textheight,clip,trim={7mm 0 0 0}]{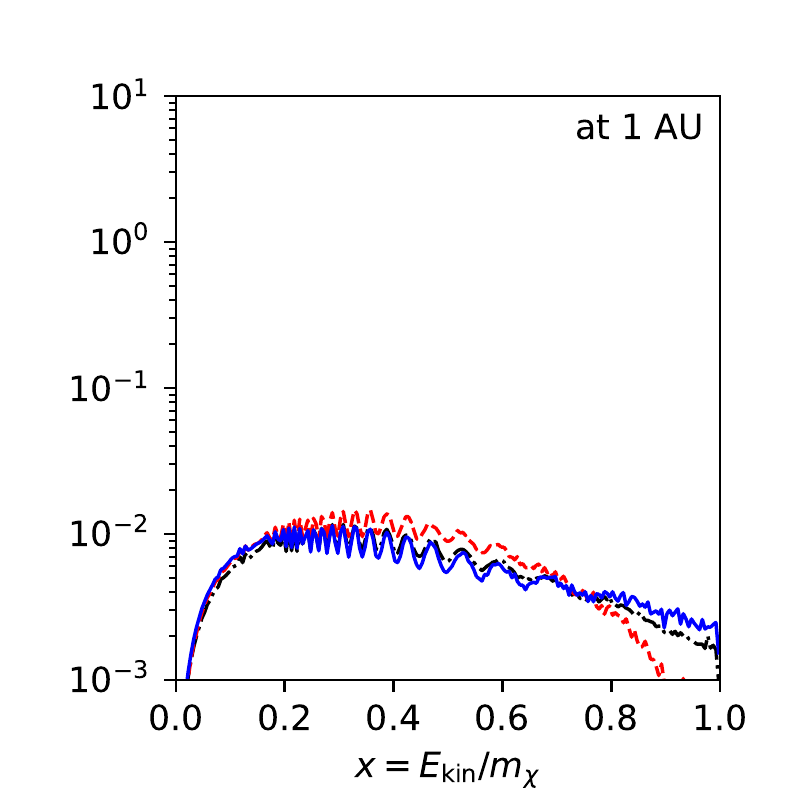} \\
    \includegraphics[height=0.3\textheight]{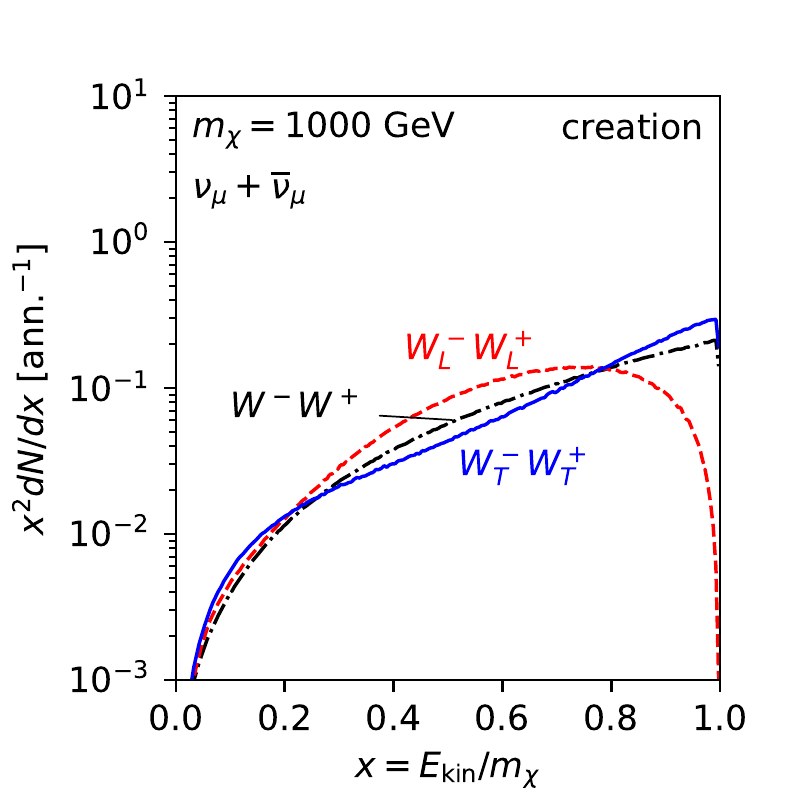}
    \includegraphics[height=0.3\textheight,clip,trim={7mm 0 0 0}]{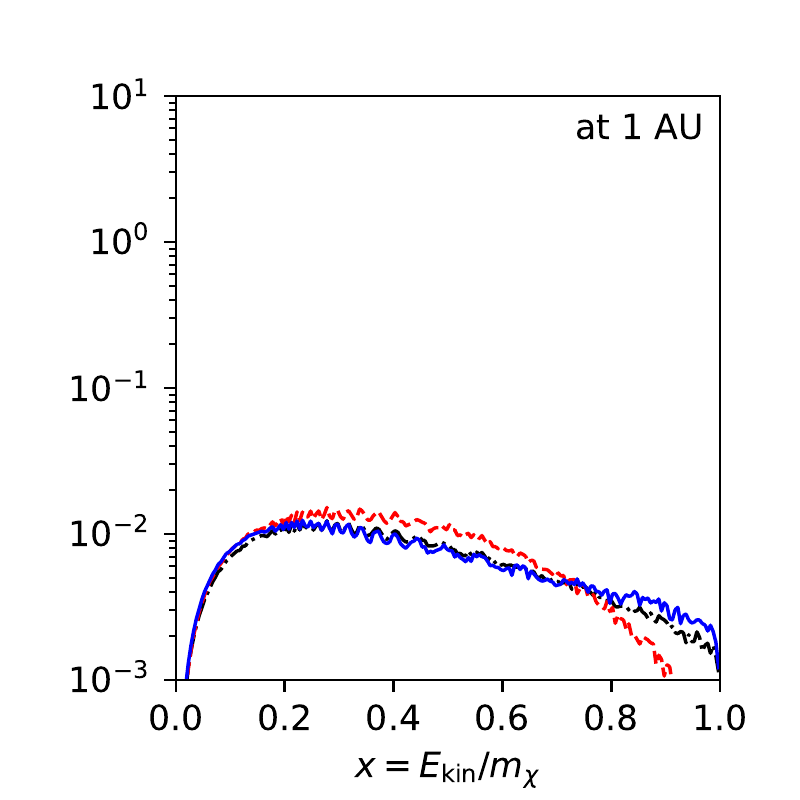} \\
    \includegraphics[height=0.3\textheight]{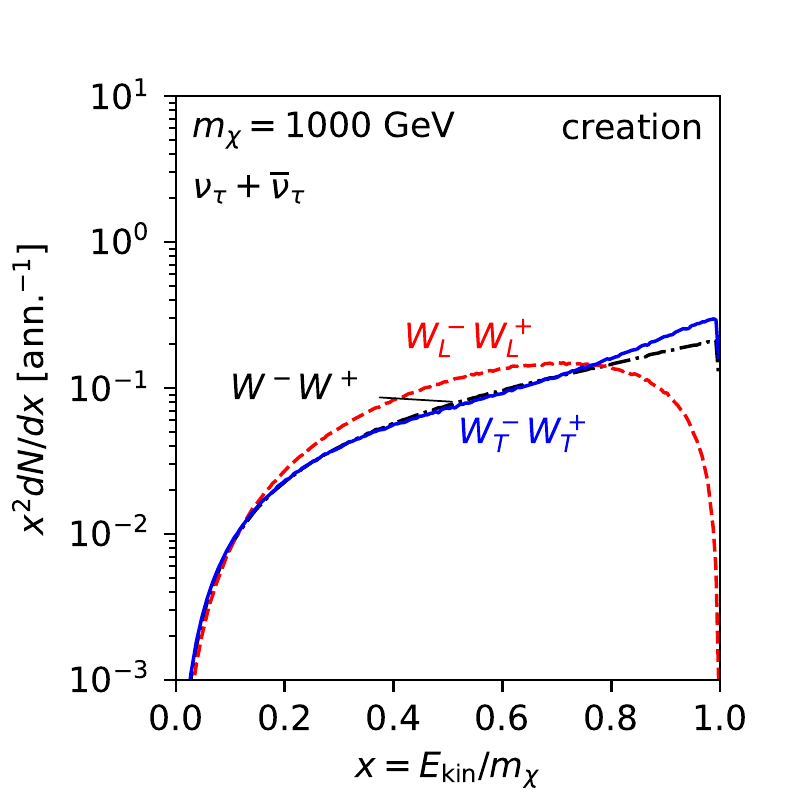}
    \includegraphics[height=0.3\textheight,clip,trim={7mm 0 0 0}]{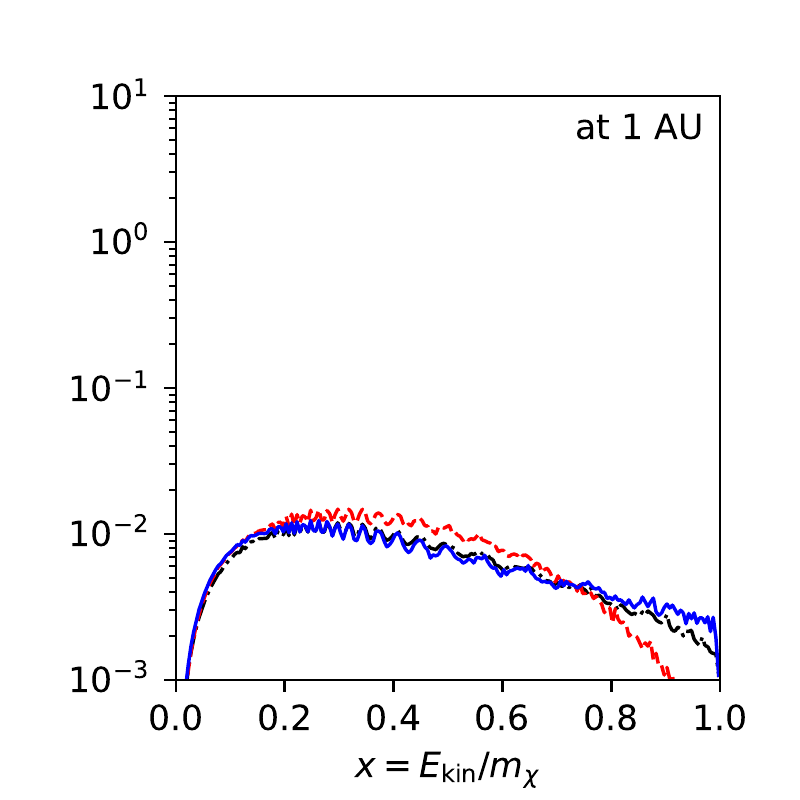}
    \caption{Same as figure~\ref{fig:ww_m100} but with a dark matter mass of \SI{1000}{\gev}. }
    \label{fig:ww_m1000}
\end{figure}

\begin{figure}[tp]
    \centering
    \includegraphics[height=0.3\textheight]{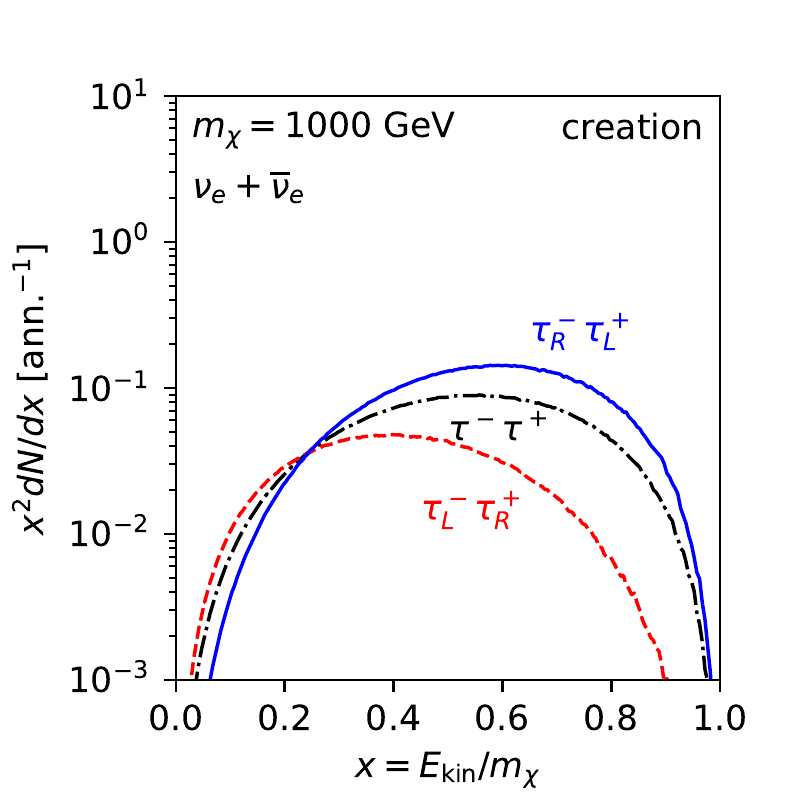}
    \includegraphics[height=0.3\textheight,clip,trim={7mm 0 0 0}]{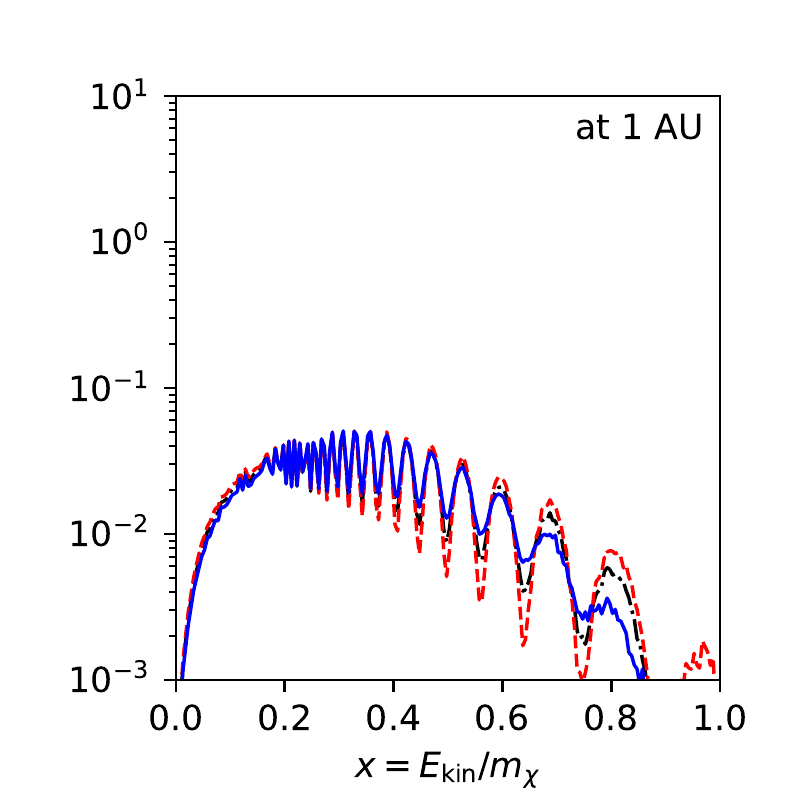} \\
    \includegraphics[height=0.3\textheight]{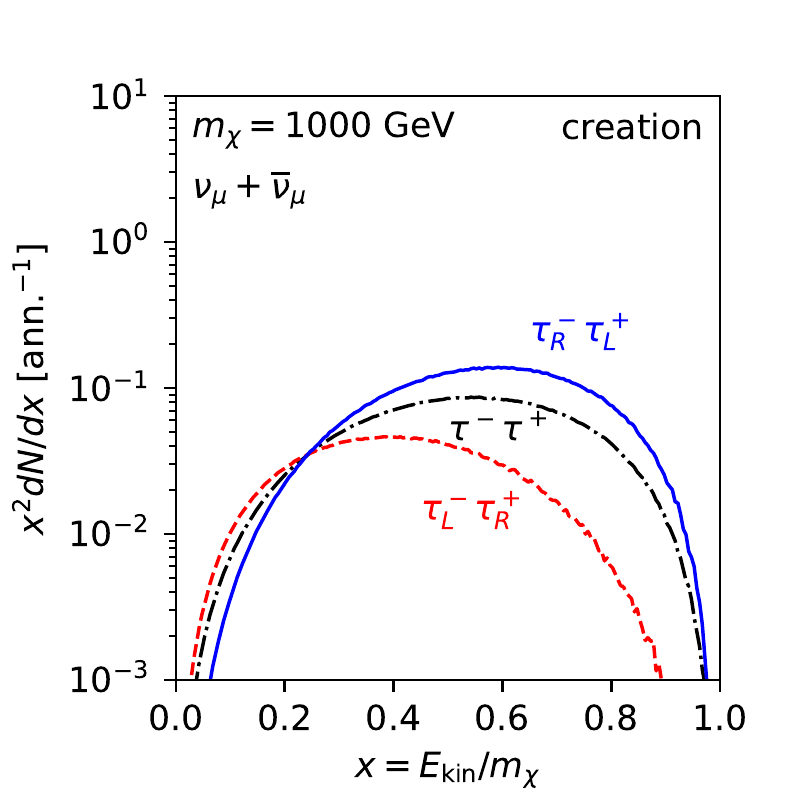}
    \includegraphics[height=0.3\textheight,clip,trim={7mm 0 0 0}]{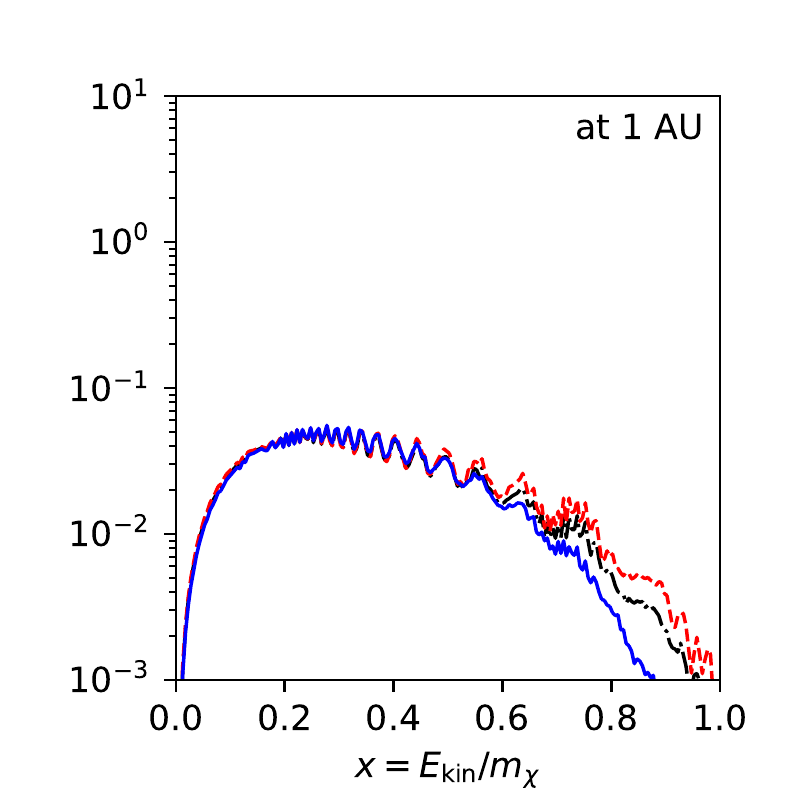} \\
    \includegraphics[height=0.3\textheight]{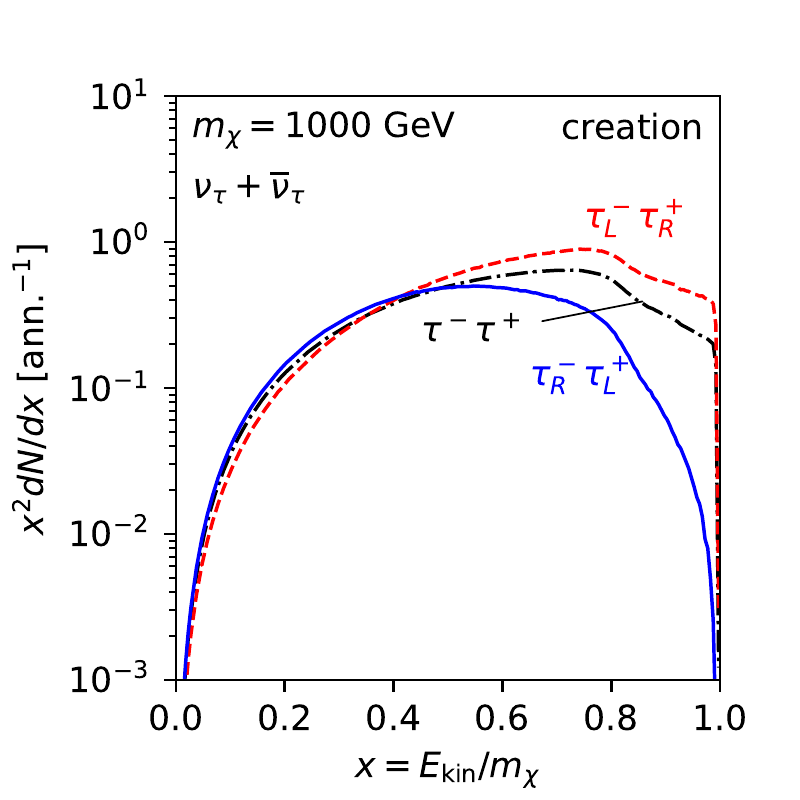}
    \includegraphics[height=0.3\textheight,clip,trim={7mm 0 0 0}]{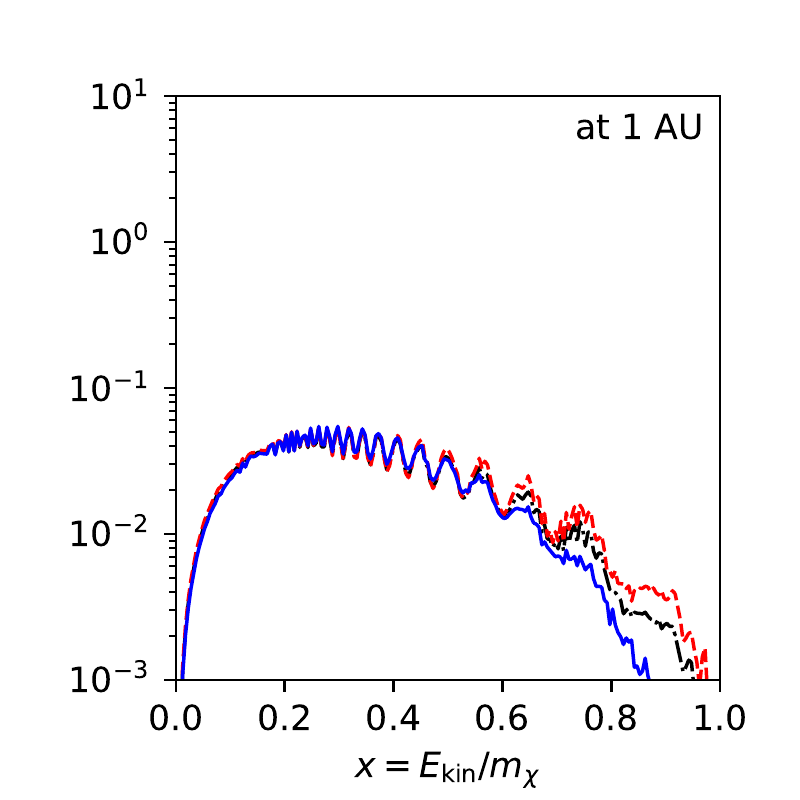}
    \caption{Same as figure~\ref{fig:tau_m100} but with a dark matter mass of \SI{1000}{\gev}. }
    \label{fig:tau_m1000}
\end{figure}

In figures~\ref{fig:ww_m100},~\ref{fig:tau_m100},~\ref{fig:ww_m1000} and~\ref{fig:tau_m1000} we show neutrino fluxes obtained in the \wimpsim simulations for the $W$ and $\tau$ annihilation channels. We show the sum of neutrino and antineutrino fluxes for all neutrino flavours, in the left column of the figures at creation and in the right column after propagation to \SI{1}{\au} from the Sun. We show in figures~\ref{fig:ww_m100} and ~\ref{fig:tau_m100} the fluxes for a dark matter mass of \SI{100}{\gev}, where interactions in the Sun are less important. To highlight the effect of interactions we show in figures~\ref{fig:ww_m1000} and ~\ref{fig:tau_m1000} the fluxes for a dark matter mass of \SI{1000}{\gev}. 

Generally, we can see the effect of oscillations between the annihilation point and the Earth as sinusoidal behaviours in the energy spectra. The oscillations will appear in given ranges of neutrino energies (with one energy range corresponding to each of the neutrino squared mass differences) and are mainly visible above \SI{100}{\gev} in the spectra in the right-hand parts of figs.~\ref{fig:ww_m1000} and~\ref{fig:tau_m1000} for $m_\chi=\SI{1000}{\gev}$. For more details regarding the oscillation effects we refer to ref.~\cite{Blennow:2007tw}. 

For the fluxes from the $\tau$ lepton channels, we can see that the $\nu_e+\overline{\nu}_e$ and $\nu_\mu+\overline{\nu}_\mu$ fluxes are very similar at production, whereas the $\nu_\tau+\overline{\nu}_\tau$ fluxes are significantly larger, especially at high energy. The explanation from this comes from the fact that the high energy part of the $\nu_e+\overline{\nu}_e$ and $\nu_\mu+\overline{\nu}_\mu$ fluxes come from the $W$ bosons in the $\tau$ decays, and the branching ratio into the two flavours is approximately equal whereas the high energy $\nu_\tau$ can come directly from the $\tau$ decays, resulting in higher average energies. We can also see that for the polarisation channel where the $\nu_\tau+\overline{\nu}_\tau$ flux is higher, the $\nu_e+\overline{\nu}_e$ and $\nu_\mu+\overline{\nu}_\mu$ flux is lower, and vice versa. 

For the $W$ channels at $m_\chi=\SI{100}{\gev}$, we can clearly see the box-shaped spectrum (which here appears tilted, since we show $x^2 dN/dx$ on the vertical axis), is now either concave, linear or convex depending on the polarisation of the $W$ bosons. In the $\nu_\tau+\overline{\nu}_\tau$ fluxes there is an additional component in the lower end of the box, coming from $\tau$ decays. 

In general, we can see in our results that differences between polarisation states at creation can be to some extent washed out by propagation effects (interactions and oscillations). For example we can see that for the $\tau$ annihilation channels, although there is a rather large difference between the $\nu_\mu+\overline{\nu}_\mu$ fluxes from the $\tau_L^- \tau_R^+$ and $\tau_R^- \tau_L^+$ channels at creation, this difference becomes smaller at \SI{1}{\au}. This is primarily due to the oscillations between $\nu_\tau$ and $\nu_\mu$, which means that the $\nu_\mu+\overline{\nu}_\mu$ flux at \SI{1}{\au} is approximately the average between the two neutrino flavours at creation, and this average is similar for the two polarisation channels (and closer to the unpolarised case). 

Interestingly, we can also see that for the muon neutrinos, the $\tau_L^- \tau_R^+$ flux is lower than the $\tau_R^- \tau_L^+$ flux at creation, but larger at \SI{1}{\au}. This can be explained by the fact that the  $\nu_\tau+\overline{\nu}_\tau$ flux is instead higher for the $\tau_L^- \tau_R^+$ channel than for the $\tau_R^- \tau_L^+$ channel, and sufficiently much larger to end up with a larger $\nu_\mu+\overline{\nu}_\mu$ flux for $\tau_L^- \tau_R^+$ at \SI{1}{\au}.

For the $W$ boson channels, it is interesting to see that the propagated flux at \SI{1}{\au} is not very different from the production flux, apart from a slight shift towards lower energies, likely caused by interactions in the Sun. Thus a difference between the polarisation states remains present at the highest neutrino energies for the $W$ channels in the \SI{1}{\au} flux. 

Apart from oscillations, propagation effects also include neutrino interactions in the Sun. For the $m_\chi=\SI{1000}{\gev}$ fluxes, we clearly see how interactions act to attenuate the fluxes and shift them towards lower energy. This effect also decreases the differences between the polarisation states. The high energy peak for the $W_T^- W_T^+$ channel in particular is damped by the interactions, making it less distinguishable from the $W_L^- W_L^+$ channel, which is less affected by this attenuation effect. As the probability to detect a neutrino in a neutrino telescope goes like $E_\nu^2$ one would naively have believed that the $W_T^- W_T^+$ channel would give a higher event rate. However, for large masses it actually goes the other way due to neutrino interactions in the Sun.

We can quantify this by looking at expected number of events in a neutrino telescope for these spectra. We do this by calculating the number of events in IceCube using their latest 3 year data analysis for dark matter searches from the Sun \cite{Aartsen:2016zhm}. We use their given effective area (we will use the maximum of DeepCore and IceCube effective areas for each given energy) and calculate the expected number of events (per annihilation) for the lifetime of the detector in these three years for our different spectra (including both muon neutrinos and antineutrinos). Our results are shown in table~\ref{tab:icrates}. As one can see, the shape of the $W_T^- W_T^+$ spectrum gives a higher event rate than the $W_L^- W_L^+$ spectrum for a 100 GeV WIMP mass. However, for large masses, neutrino interactions in the Sun cause this harder spectrum to be absorbed more, and the end result is that the event rate for a 1000 GeV WIMP is actually lower than for the $W_L^- W_L^+$ spectrum. The same trend can be seen for the $\tau_R^- \tau_L^+$ spectra compared to the $\tau_L^- \tau_R^+$ spectra, even though in this case, the $\tau_R^- \tau_L^+$ spectrum still gives a higher event rate even for a 1000 GeV WIMP. The unpolarised \pythias results are typically in between our new polarised results, except for $W^-W^+$ at 1000 GeV, where the $\pythias$ results is comparable to the $W_T^- W_T^+$ results (which can also be seen from the spectra in figure~\ref{fig:ww_m1000}).

\renewcommand{\arraystretch}{1.3}
\begin{table}[t]
    \centering
    \begin{tabular}{l c ccc c}\hline\hline
        \multirow{2}{*}{${\boldsymbol m_\chi}$ [GeV]} & \multirow{2}{*}{{\bfseries Channel}} &  \multicolumn{3}{c}{ \bfseries Events per annihilation}  & {\bfseries Ratio} \\
        & & \ps (unpol.) & $W_L^- W_L^+$ & $W_T^- W_T^+$ &  (col 5/col 4) \\ \hline 
         100 & \multirow{2}{*}{$W^-W^+$} & \num{8.99e-25}  & \num{8.70e-25}  & \num{9.58e-25}  & 1.13  \\
         1000 & & \num{3.44e-23}  & \num{4.04e-23}  & \num{3.46e-23} & 0.86 \\ \hline  
        & & \ps (unpol.) & $\tau_R^- \tau_L^+$ & $\tau_L^- \tau_R^+$ & \\ \hline
         100 & \multirow{2}{*}{$\tau^-\tau^+$} & \num{2.13e-24}  & \num{1.66e-24}  & \num{2.60e-24}  & 1.56 \\ 
         1000 &  & \num{1.16e-22}  & \num{1.14e-22}  & \num{1.24e-22} & 1.08 \\ \hline\hline
    \end{tabular}
    \caption{The number of events (per annihilation) in IceCube for different masses, annihilation channels and polarisation states. The \pythias (\ps) results are from a regular \wimpsim run and the results for the polarised states are for the runs specified in this section. The IceCube number of events are calculated with the effective areas in ref.~\cite{Aartsen:2016zhm} for a livetime of 532 days (corresponds to three years of data taking). The ratio in the last column is the ratio of the two polarisation results (column 5/column 4).}
    \label{tab:icrates}
\end{table}
\renewcommand{\arraystretch}{1}

We end this section by noting that the differences we see for the $W^-W^+$ final state polarisations are slightly more than the 5--10\% quoted by \cite{Jungman:1994jr}. For the $\tau^-\tau^+$ channel, the effect can be even larger than this. However, for more massive dark matter, interactions tend to wash out some of the differences, even though some differences remain. This effect of washing out differences for more massive dark matter has also been observed in ref.~\cite{Cirelli:2005gh,Barger:2007xf}

\section{Summary and conclusions} \label{sec:conclusions}
In this study we have investigated the fluxes of secondary particles from dark matter annihilations, such as neutrinos, charged antiparticles and gamma-rays. A proper knowledge of these fluxes is necessary in order to analyse results in indirect searches for dark matter. We have simulated the dark matter annihilations with two of the most commonly used event generators, \pythiae and \herwigs (in combination with \madgraph), in order to understand the impact that the differences between the modelling in the generators can have.  For some of the steps in the simulation chain, the event generators rely on phenomenological modelling of the underlying physics, and this modelling in general differs between generators. In particular this concerns the QCD-related parts of the process, which is most relevant when the dark matter annihilation proceeds into colour charged final state particles such as quarks.

We have also looked at the effect that polarisation of the final state can have on the fluxes of the secondary particles. Polarisation of the final state is expected in many concrete dark matter models, for example if the annihilation proceeds through a chiral coupling or if there is a separation between the couplings to the longitudinal and transverse degrees of freedom of gauge bosons. 

Concerning the comparison between \pythiae and \herwigs, we find the smallest differences in the fluxes of neutrinos and positrons and the largest differences for antiprotons and, in some cases, gamma-rays, with the spectra varying by a factor of 2 or more for some energies. Differences in the leptonic spectra are typically of the order of 10-20\% or lower. Apart from the gamma-ray fluxes, where a difference in the modelling of QED bremsstrahlung is a likely cause of the differences, it is expected that antiproton fluxes will show larger discrepancies than leptonic fluxes, since the antiprotons emerge from the hadronisation part of the event generation, which the event generators model differently.  While uncertainties in the yields from event generators are often ignored in indirect detection analyses, these results illustrate the need for these uncertainties to be included, particularly for the case of hadronic processes.

The polarisation of the final state instead primarily affects the leptonic spectra. This is because the most energetic leptons are the ones coming from decays early in the chain, close to the hard process, whereas in the showering and hadronisation processes, producing the antiprotons and many of the gamma-rays, polarisation effects are washed out. Differences in the leptonic spectra for the different polarisations of the final state in many cases reaches a factor of ten, demonstrating the importance of taking into account the polarisations predicted by a model in determining these yields. The gamma-rays and antiprotons are not produced directly as decay products of the final states, and differences in these spectra are most likely traced to spectral changes for the other particles involved. 

We have also looked more specifically at the neutrino fluxes from dark matter annihilations in the Sun. For this purpose, it is important to include the interactions of neutrinos with the solar material and neutrino oscillations when propagating the neutrinos from the annihilation point in the solar core to a neutrino telescope on Earth. We use the code \wimpsim to account for these effects. 

Interestingly, although the differences in the neutrino spectra for different polarisation states are relatively large at production in the annihilation process, these differences are to an extent reduced by the propagation effects. For example, for annihilations into $\tau$ leptons, in the polarisation state where the muon neutrino flux is larger, the tau neutrino flux is lower, and vice versa. Since these neutrino flavours mix significantly in the oscillations, the general size of the resulting muon neutrino flux at Earth ends up in between the fluxes of the two flavours at production, and this flux is similar for the two polarisation states. Hence, the polarisation differences are ultimately averaged out by the oscillations. 

Neutrino interactions in the Sun also act to reduce the differences between the polarisation states. We have shown that for larger values of the dark matter mass, the attenuation of the fluxes at high energy, stemming from charged current interactions of the neutrinos with the solar material, dampens the high energy peak in the neutrino spectrum from transversely polarised $W$ bosons in the final state. This dampening reduces the differences in the neutrino flux from the $W_T^- W_T^+$ and $W_L^- W_L^+$ channels. At lower dark matter masses, where interactions are less important, there is however still a difference between the polarisations remaining in the propagated neutrino fluxes.

Finally, to quantify the above, we have looked at the differences in the event rates in IceCube. We find that the different polarisation states have an effect on the event rates which is typically around 10\% in the cases considered but reaches over 50\% in the $\tau$ annihilation channel at \SI{100}{\gev}.

\acknowledgments
The authors wish to thank Philip Ilten for assistance with $\tau$ lepton decays in \pythia and Marco Cirelli and Jared Evans for helpful discussions and correspondence. JMC gratefully acknowledges partial support from US Department of Energy (DOE) Grant No. DE-SC0020047 and expresses a special thanks to the Mainz Institute of Theoretical Physics (MITP), which is part of the DFG Cluster of Excellence PRISMA* (Project ID 39083149), for its hospitality and support. JE thanks the Swedish Research Council for support (contract 621-2014-5772).


\bibliographystyle{JHEP}
\bibliography{dmann}

\clearpage


\begin{appendix}
\section{Yields for $m_\chi=\SI{100}{\gev}$}
\label{app:100GeV}
In this section we show figures similar to figures~\ref{fig:yields_nu_hvsp}-\ref{fig:yields_pbar_hvsp}, but for a dark matter mass of \SI{100}{\gev}. 
\begin{figure}[htb]
    \centering
    \includegraphics[height=0.26\textheight,clip,trim={0 0 0 8mm}]{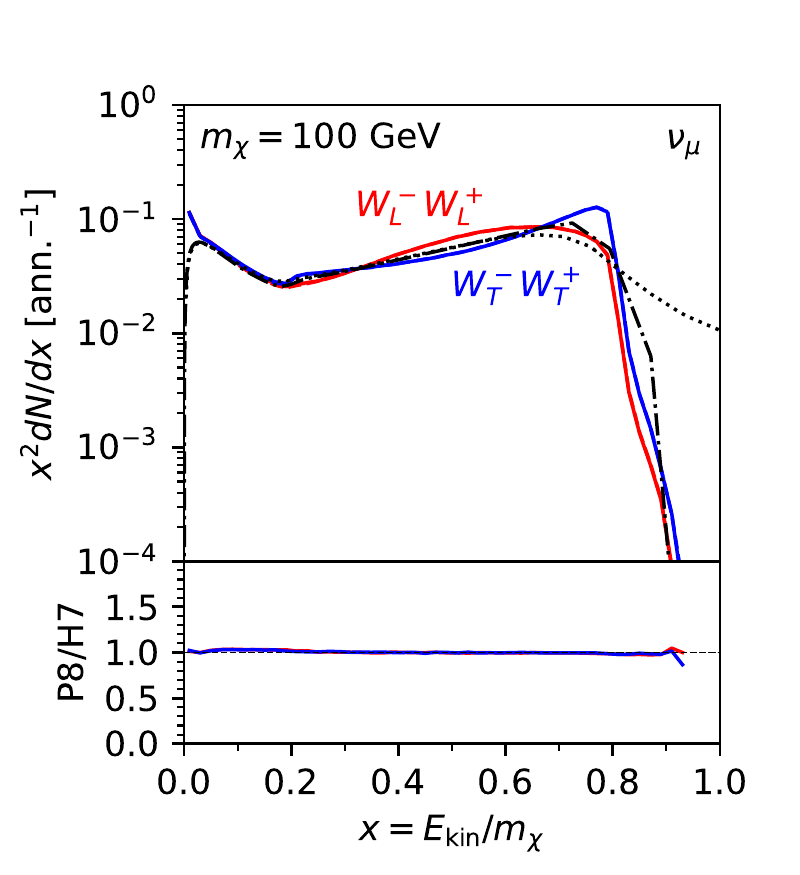}\hspace{-1.0em}
    \includegraphics[height=0.26\textheight,clip,trim={14mm 0 0 8mm}]{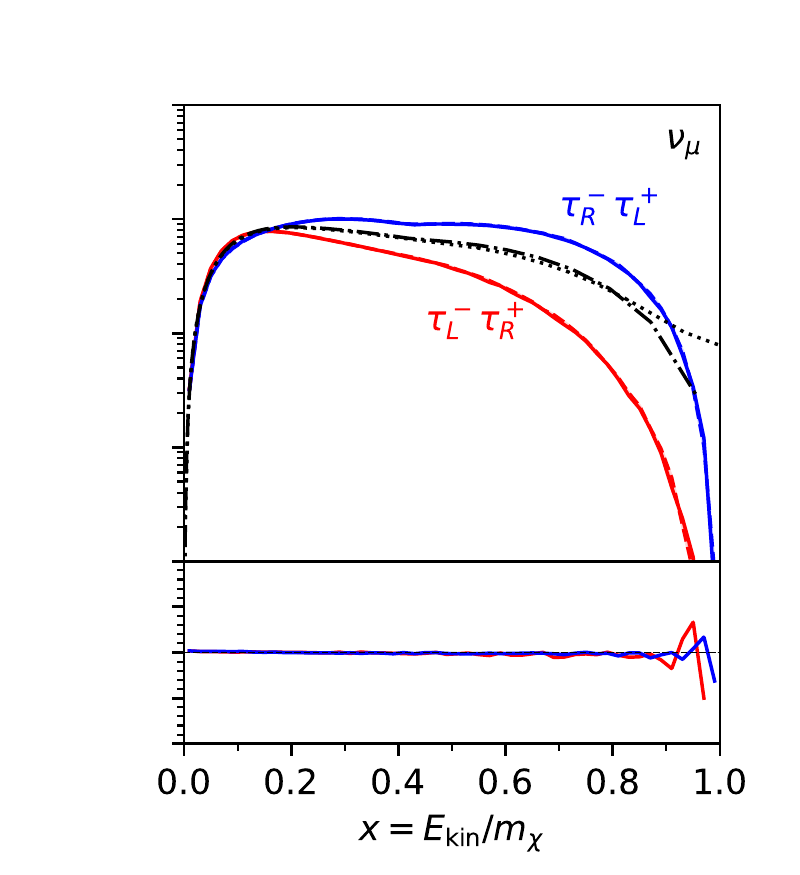}\hspace{-1.0em}
    \includegraphics[height=0.26\textheight,clip,trim={14mm 0 0 8mm}]{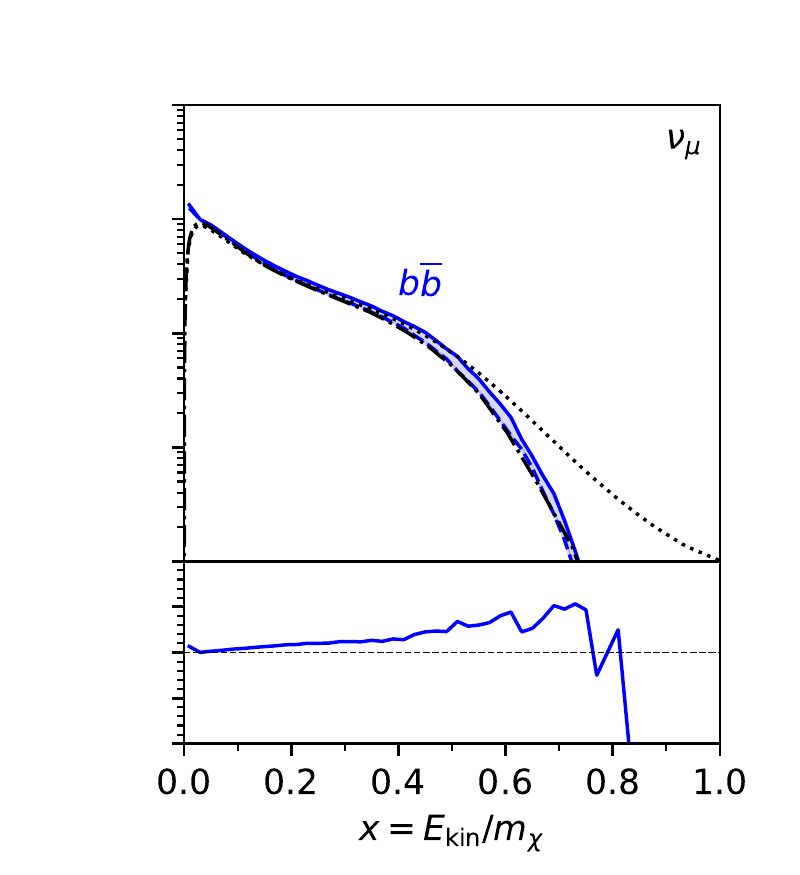}
    \caption{Yields of muon neutrinos, comparing simulations done with \pythiae (solid) with simulations done with \herwigs (dashed). We also show the spectra for unpolarised final states simulated with \pythias (extracted from \darksusy) in the dash-dotted curves and the results of the \pppc \cite{Cirelli:2010xx} with dotted curves. The annihilation channels are indicated in each panel. The dark matter mass is here set to $m_\chi=\SI{100}{\gev}$. In the plots, the upper part show $x^2dN/dx$ normalised to the number of simulated annihilations on the vertical axis whereas the lower part of each plot show the ratio, defined as $\text{\pe}/\text{\hs}$. The horisontal axis shows $x=E_{\rm kin}/m_\chi$ in all plots. }
    \label{fig:yields_nu_hvsp_100}
\end{figure}
\begin{figure}
    \centering
    \includegraphics[height=0.26\textheight,clip,trim={0 0 0 8mm}]{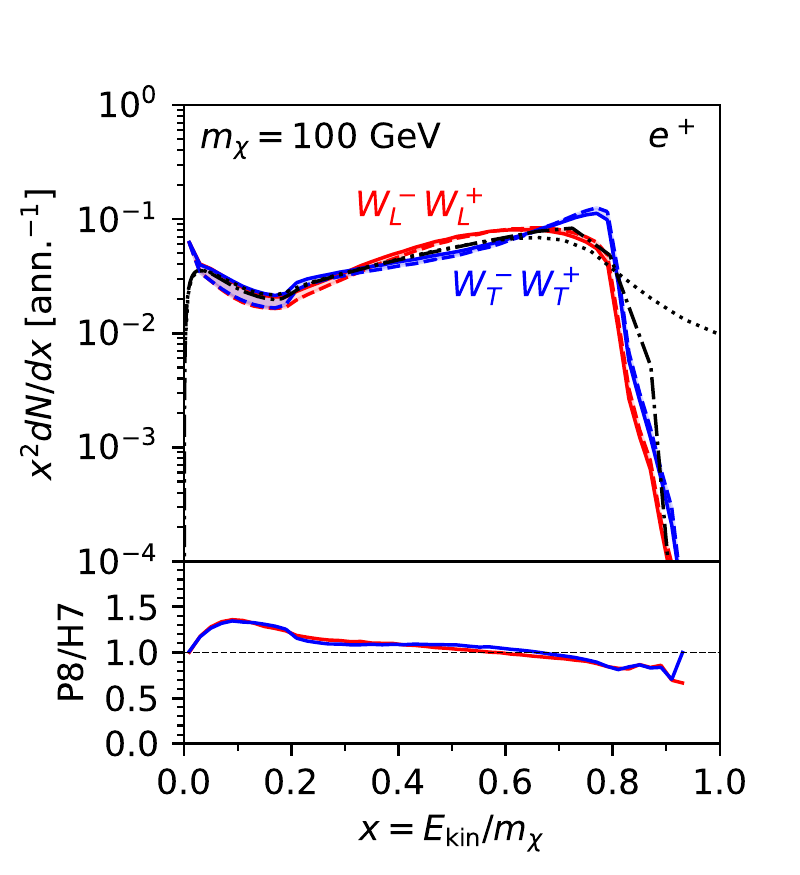}\hspace{-1.0em}
    \includegraphics[height=0.26\textheight,clip,trim={14mm 0 0 8mm}]{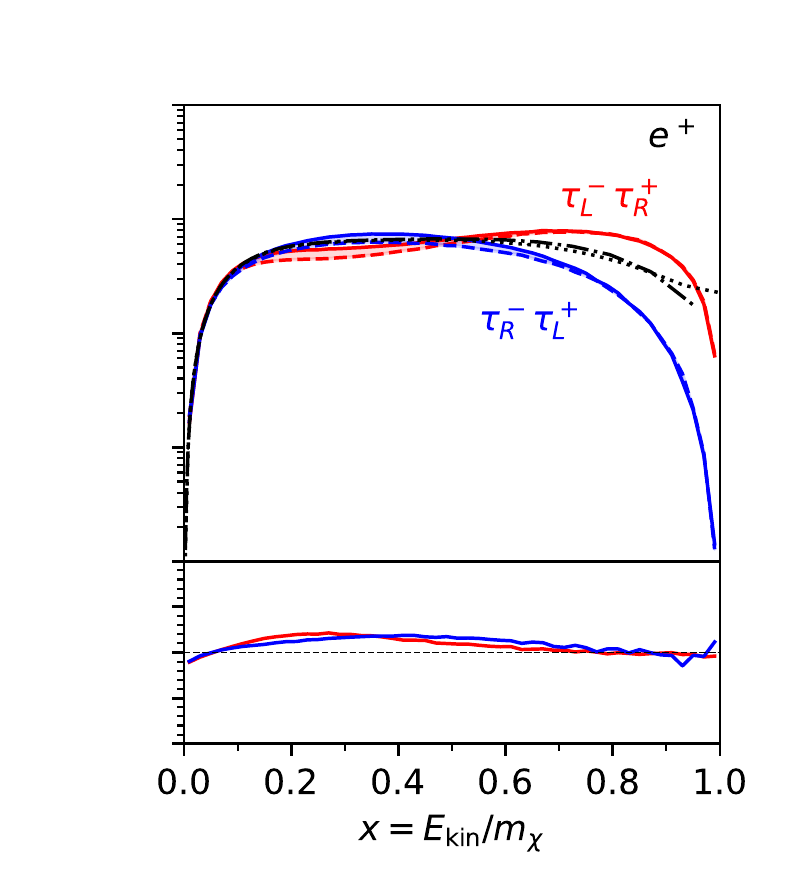} \hspace{-1.0em}
    \includegraphics[height=0.26\textheight,clip,trim={14mm 0 0 8mm}]{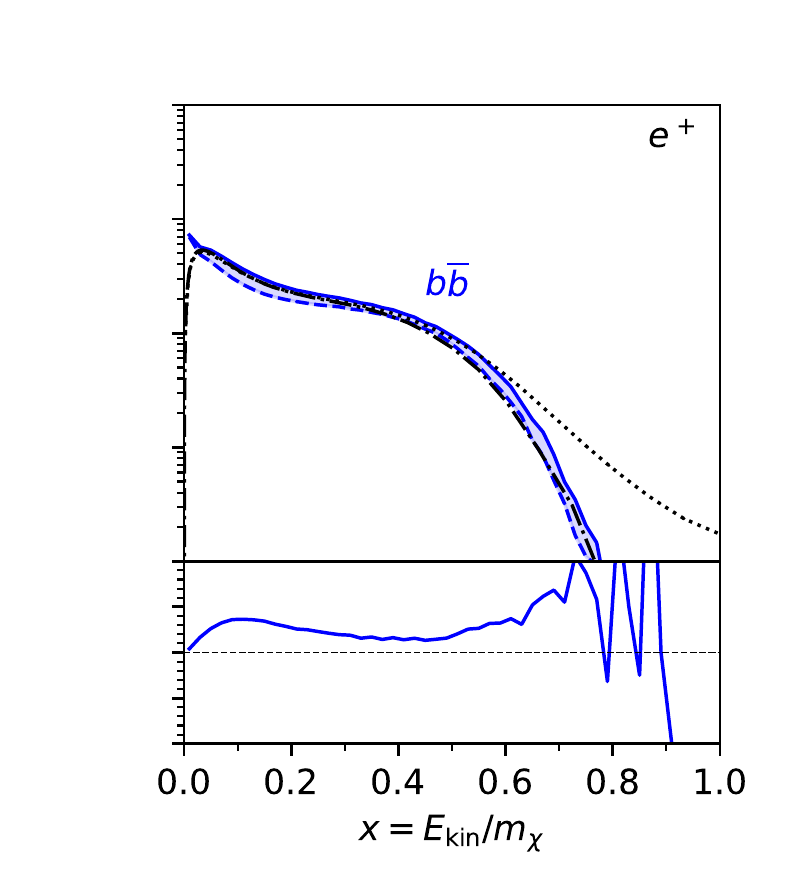}
    \caption{Same as figure~\ref{fig:yields_nu_hvsp_100} but for positron yields.}
    \label{fig:yields_ep_hvsp_100}
\end{figure}
\begin{figure}[t]
    \centering
    \includegraphics[height=0.26\textheight,clip,trim={0 0 0 8mm}]{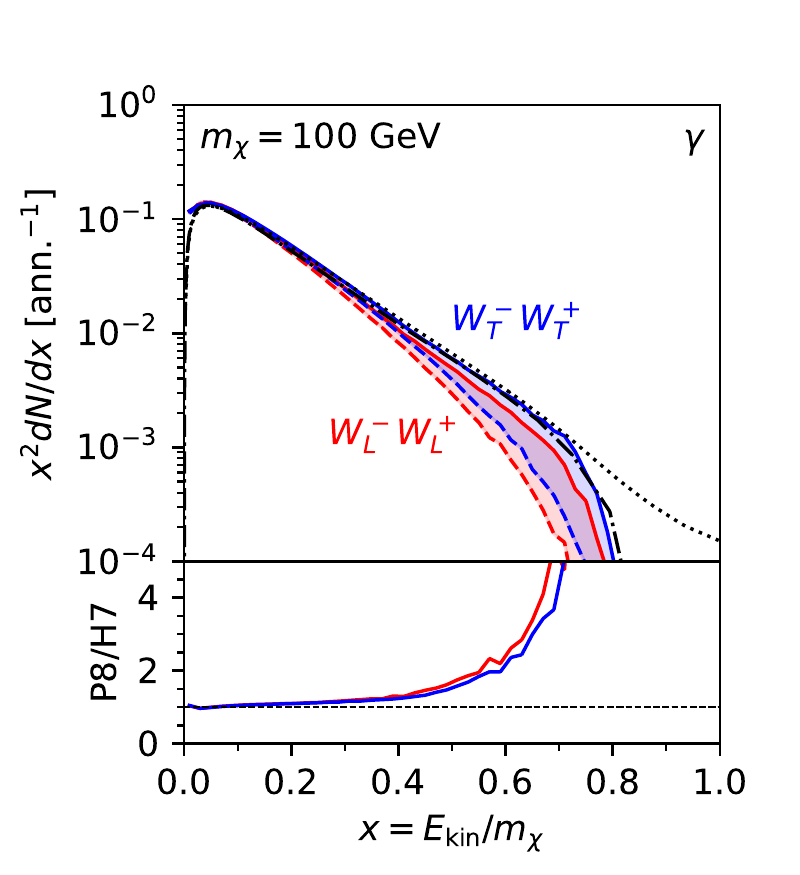}\hspace{-1.0em}
    \includegraphics[height=0.26\textheight,clip,trim={14mm 0 0 8mm}]{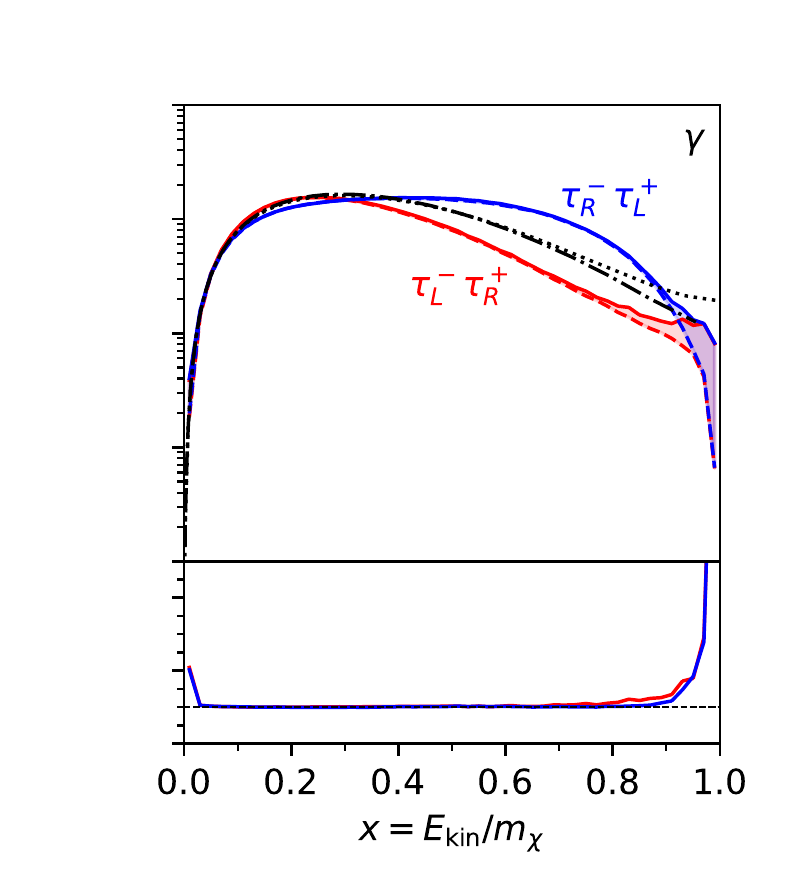}\hspace{-1.0em}
    \includegraphics[height=0.26\textheight,clip,trim={14mm 0 0 8mm}]{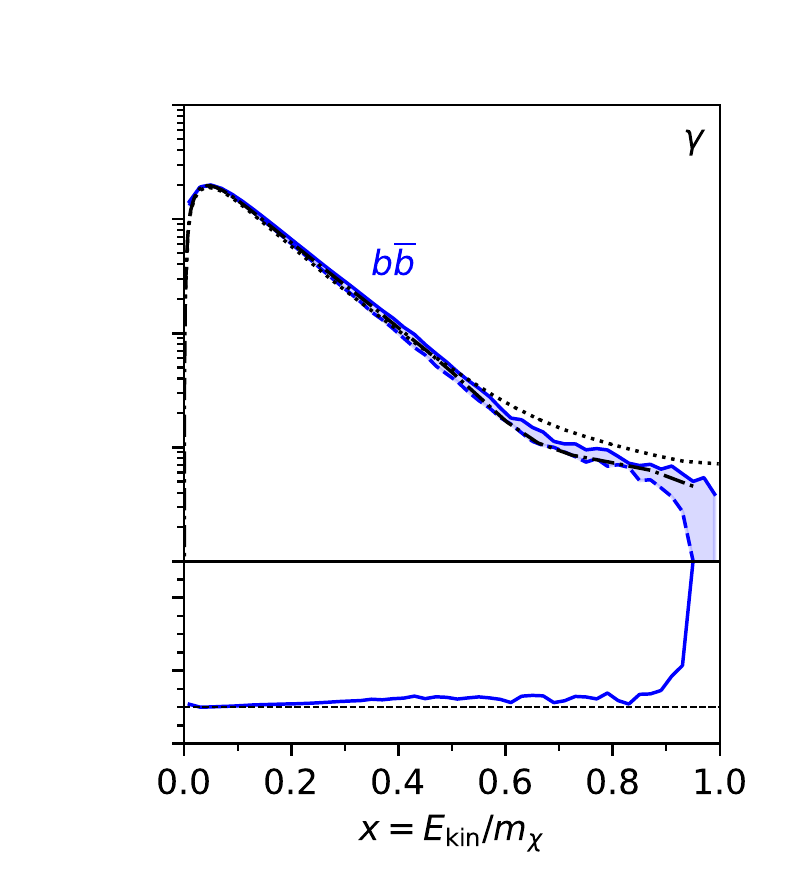}
    \caption{Same as figure~\ref{fig:yields_nu_hvsp_100} but for gamma-ray yields.}
    \label{fig:yields_gamma_hvsp_100}
\end{figure}
\begin{figure}
    \centering
    \includegraphics[height=0.26\textheight,clip,trim={0 0 0 8mm}]{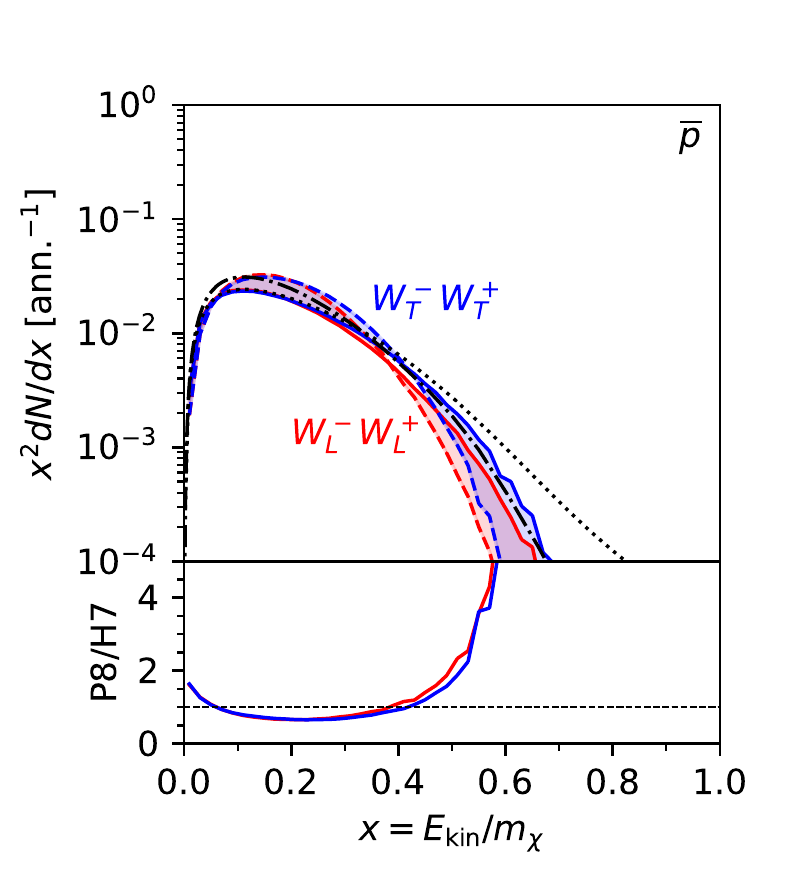}\hspace{-1.0em}
    \includegraphics[height=0.26\textheight,clip,trim={14mm 0 0 8mm}]{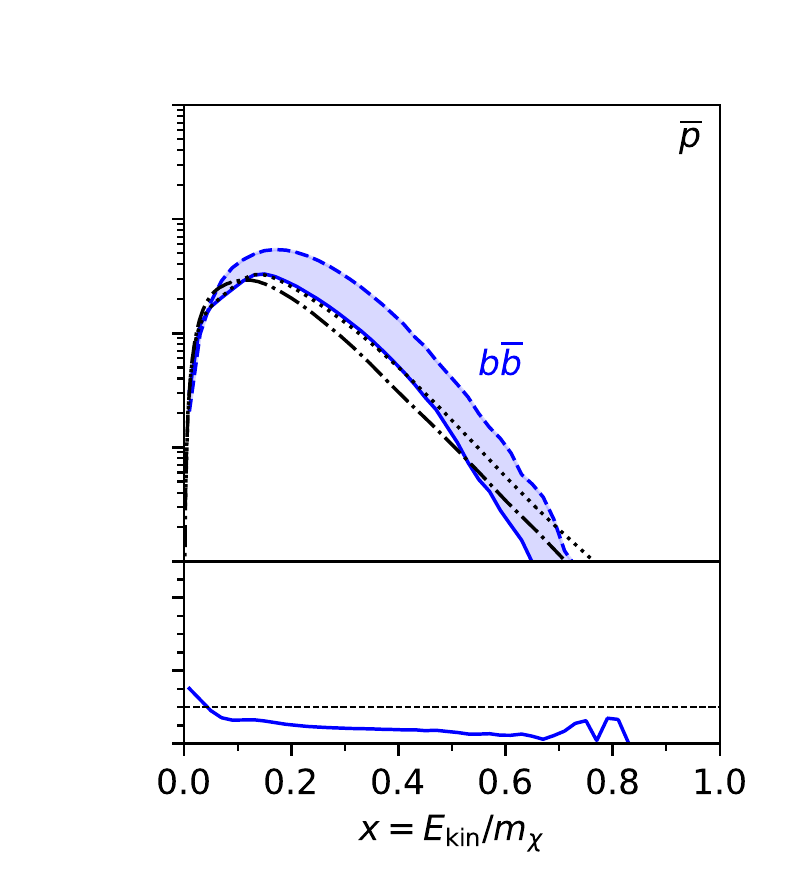}    
    \caption{Same as figure~\ref{fig:yields_nu_hvsp_100} but for antiproton yields.}
    \label{fig:yields_pbar_hvsp_100}
\end{figure}

\end{appendix}

\end{document}